\documentclass[amssymb,amsmath,prd,twocolumn,preprintnumbers,superscriptaddress,nofootinbib]{revtex4-1}

\usepackage[10pt]{type1ec}  
\usepackage[T1]{fontenc}
\usepackage{CJKutf8}
\usepackage[overlap, CJK]{ruby}
\usepackage{CJKulem}
\newenvironment{SChinese}{%
  \CJKfamily{gbsn}%
  \CJKtilde
  \CJKnospace}{}

\usepackage{graphicx}
\usepackage{bm}
\usepackage{amssymb,amsmath}
\usepackage{mathrsfs}
\usepackage{latexsym}
\usepackage{color}
\usepackage{dcolumn}
\usepackage[colorlinks=true,citecolor=blue,urlcolor=blue]{hyperref}
\usepackage[usenames,dvipsnames]{xcolor}
\usepackage{xspace}
\usepackage{graphicx}

\usepackage{multirow}
\usepackage{xfp} 

\newcommand{\externalresult}[1]{\textcolor{black}{#1}}
\newcommand{\result}[1]{\textcolor{black}{#1}}

\newcommand{\PnucCoef}{\ensuremath{10^{33}}}
\newcommand{\PtwonucCoef}{\ensuremath{10^{34}}}
\newcommand{\PsixnucCoef}{\ensuremath{10^{35}}}

\newcommand{\PriorMmax  }{\result{\ensuremath{1.47^{+0.71}_{-1.37}}}} 
\newcommand{\PriorDeltaR}{\result{\ensuremath{0.48^{+1.28}_{-6.67}}}} 
\newcommand{\PriorROnePointFour}{\result{\ensuremath{8.09^{+5.68}_{-3.96}}}} 

\newcommand{\PriorRTwoPointZero}{\result{\ensuremath{5.90^{+6.97}_{-0.00}}}} 

\newcommand{\PriorLOnePointFour}{\result{\ensuremath{24^{+841}_{-24}}}} 

\newcommand{\PriorLTwoPointZero}{\result{\ensuremath{0^{+54}_{-0}}}} 

\newcommand{\PriorrhocOnePointFour}{\result{\ensuremath{8.4^{+12.5}_{-6.0}}}} 
\newcommand{\PriorrhocTwoPointZero}{\result{\ensuremath{9.0^{+5.7}_{-6.3}}}} 
\newcommand{\PriorrhocMmax        }{\result{\ensuremath{2.4^{+0.9}_{-2.0}}}} 

\newcommand{\PriorPnuc   }{\result{\ensuremath{2.25^{+5.81}_{-2.15}}}} 
\newcommand{\PriorPtwonuc}{\result{\ensuremath{1.22^{+4.86}_{-1.21}}}} 
\newcommand{\PriorPsixnuc}{\result{\ensuremath{2.43^{+4.70}_{-2.43}}}} 

\newcommand{\PriorMaxCs   }{\result{\ensuremath{0.76^{+0.24}_{-0.37}}}} 
\newcommand{\PriorRhoMaxCs}{\result{\ensuremath{1.38^{+1.65}_{-1.34}}}} 
\newcommand{\PriorPMaxCs  }{\result{\ensuremath{1.65^{+8.16}_{-1.65}}}} 

\newcommand{\PSRMmax  }{\result{\ensuremath{2.24^{+0.48}_{-0.24}}}} 
\newcommand{\PSRDeltaR}{\result{\ensuremath{-0.07^{+1.00}_{-1.04}}}} 

\newcommand{\PSRROnePointFour}{\result{\ensuremath{13.54^{+2.61}_{-3.13}}}} 

\newcommand{\PSRRTwoPointZero}{\result{\ensuremath{13.18^{+3.02}_{-2.90}}}} 

\newcommand{\PSRLOnePointFour}{\result{\ensuremath{795^{+1262}_{-708}}}} 

\newcommand{\PSRLTwoPointZero}{\result{\ensuremath{66^{+184}_{-66}}}} 

\newcommand{\PSRrhocOnePointFour}{\result{\ensuremath{5.7^{+3.2}_{-3.1}}}} 
\newcommand{\PSRrhocTwoPointZero}{\result{\ensuremath{8.5^{+4.8}_{-5.3}}}} 
\newcommand{\PSRrhocMmax        }{\result{\ensuremath{1.4^{+0.5}_{-0.6}}}} 

\newcommand{\PSRPnuc   }{\result{\ensuremath{6.07^{+7.53}_{-5.93}}}} 
\newcommand{\PSRPtwonuc}{\result{\ensuremath{6.00^{+4.79}_{-5.99}}}} 
\newcommand{\PSRPsixnuc}{\result{\ensuremath{7.51^{+6.77}_{-5.15}}}} 

\newcommand{\PSRMaxCs   }{\result{\ensuremath{0.72^{+0.28}_{-0.26}}}} 
\newcommand{\PSRRhoMaxCs}{\result{\ensuremath{0.97^{+0.64}_{-0.70}}}} 
\newcommand{\PSRPMaxCs  }{\result{\ensuremath{2.68^{+5.18}_{-2.68}}}} 

\newcommand{\OldMmax  }{\result{\ensuremath{2.20^{+0.30}_{-0.19}}}} 
\newcommand{\OldDeltaR}{\result{\ensuremath{-0.17^{+0.85}_{-0.83}}}} 

\newcommand{\OldROnePointFour}{\result{\ensuremath{12.25^{+1.13}_{-1.33}}}} 
\newcommand{\OldROnePointFourLow}{10.92}

\newcommand{\OldRTwoPointZero}{\result{\ensuremath{12.05^{+1.18}_{-1.45}}}} 
\newcommand{\OldRTwoPointZeroLow}{10.60}

\newcommand{\OldLOnePointFour}{\result{\ensuremath{442^{+235}_{-274}}}} 
\newcommand{\OldLOnePointFourLow}{168}

\newcommand{\OldLTwoPointZero}{\result{\ensuremath{34^{+35}_{-27}}}} 
\newcommand{\OldLTwoPointZeroLow}{7}

\newcommand{\OldrhocOnePointFour}{\result{\ensuremath{7.2^{+2.6}_{-1.7}}}} 
\newcommand{\OldrhocTwoPointZero}{\result{\ensuremath{10.5^{+4.1}_{-3.8}}}} 
\newcommand{\OldrhocMmax        }{\result{\ensuremath{1.6^{+0.3}_{-0.4}}}} 

\newcommand{\OldPnuc   }{\result{\ensuremath{4.05^{+3.59}_{-3.74}}}} 
\newcommand{\OldPtwonuc}{\result{\ensuremath{3.75^{+2.36}_{-2.98}}}} 
\newcommand{\OldPsixnuc}{\result{\ensuremath{8.33^{+5.22}_{-4.14}}}} 

\newcommand{\OldMaxCs   }{\result{\ensuremath{0.84^{+0.16}_{-0.28}}}} 
\newcommand{\OldRhoMaxCs}{\result{\ensuremath{1.13^{+0.64}_{-0.63}}}} 
\newcommand{\OldPMaxCs  }{\result{\ensuremath{3.52^{+6.90}_{-3.48}}}} 

\newcommand{\NewMmax  }{\result{\ensuremath{2.21^{+0.31}_{-0.21}}}} %
\newcommand{\NewDeltaR}{\result{\ensuremath{-0.12^{+0.83}_{-0.85}}}} 

\newcommand{\NewROnePointFour}{\result{\ensuremath{12.56^{+1.00}_{-1.07}}}} 
\newcommand{\NewROnePointFourLow}{11.49}

\newcommand{\NewRTwoPointZero}{\result{\ensuremath{12.41^{+1.00}_{-1.10}}}} 
\newcommand{\NewRTwoPointZeroLow}{11.31}

\newcommand{\NewLOnePointFour}{\result{\ensuremath{507^{+234}_{-242}}}} 
\newcommand{\NewLOnePointFourLow}{265}

\newcommand{\NewLTwoPointZero}{\result{\ensuremath{44^{+34}_{-30}}}} 
\newcommand{\NewLTwoPointZeroLow}{14}

\newcommand{\NewrhocOnePointFour}{\result{\ensuremath{6.7^{+1.7}_{-1.3}}}} 
\newcommand{\NewrhocTwoPointZero}{\result{\ensuremath{9.7^{+3.6}_{-3.1}}}} 
\newcommand{\NewrhocMmax        }{\result{\ensuremath{1.5^{+0.3}_{-0.4}}}} 
\newcommand{\NewrhocMmaxsat     }{\result{\ensuremath{5.4^{+1.1}_{-1.4}}}} 

\newcommand{\NewPnuc   }{\result{\ensuremath{4.30^{+3.37}_{-3.80}}}} 
\newcommand{\NewPtwonuc}{\result{\ensuremath{4.38^{+2.46}_{-2.96}}}} 
\newcommand{\NewPsixnuc}{\result{\ensuremath{7.41^{+5.87}_{-4.18}}}} 

\newcommand{\NewMaxCs   }{\result{\ensuremath{0.75^{+0.25}_{-0.24}}}} 
\newcommand{\NewRhoMaxCs}{\result{\ensuremath{1.01^{+0.63}_{-0.53}}}} 
\newcommand{\NewRhoMaxCsText}{\result{\ensuremath{1.01^{+6.3}_{-5.3}}}} 
\newcommand{\NewPMaxCs  }{\result{\ensuremath{2.77^{+5.81}_{-2.70}}}} 

\newcommand{\NewRhoMaxCssat}{\result{\ensuremath{3.60^{+2.25}_{-1.89}}}} 

\newcommand{\NewMmaxAlt  }{\result{\ensuremath{2.19^{+0.27}_{-0.19}}}} 
\newcommand{\NewDeltaRAlt}{\result{\ensuremath{-0.20^{+0.82}_{-0.88}}}} 

\newcommand{\NewROnePointFourAlt}{\result{\ensuremath{12.34^{+1.01}_{-1.25}}}} 
\newcommand{\NewROnePointFourLowAlt}{11.09}

\newcommand{\NewRTwoPointZeroAlt}{\result{\ensuremath{12.09^{+1.07}_{-1.17}}}} 
\newcommand{\NewRTwoPointZeroLowAlt}{11.02}

\newcommand{\NewLOnePointFourAlt}{\result{\ensuremath{457^{+219}_{-256}}}} 

\newcommand{\NewLTwoPointZeroAlt}{\result{\ensuremath{35^{+32}_{-24}}}} 

\newcommand{\NewrhocOnePointFourAlt}{\result{\ensuremath{7.1^{+2.1}_{-1.5}}}} 
\newcommand{\NewrhocTwoPointZeroAlt}{\result{\ensuremath{10.4^{+3.6}_{-3.5}}}} 
\newcommand{\NewrhocMmaxAlt        }{\result{\ensuremath{1.6^{+0.3}_{-0.3}}}} 

\newcommand{\NewPnucAlt   }{\result{\ensuremath{4.15^{+3.50}_{-3.76}}}} 
\newcommand{\NewPtwonucAlt}{\result{\ensuremath{3.90^{+2.11}_{-2.88}}}} 
\newcommand{\NewPsixnucAlt}{\result{\ensuremath{7.82^{+5.47}_{-3.53}}}} 

\newcommand{\NewMaxCsAlt   }{\result{\ensuremath{0.80^{+0.20}_{-0.26}}}} 
\newcommand{\NewRhoMaxCsAlt}{\result{\ensuremath{1.10^{+0.63}_{-0.58}}}} 
\newcommand{\NewPMaxCsAlt  }{\result{\ensuremath{3.26^{+6.51}_{-3.15}}}} 

\newcommand{\BFbranches}{\ensuremath{\mathcal{B}^{n>1}_{n=1}}}

\newcommand{\LikeRatiobranches}{\ensuremath{\max \mathcal{L}^{n>1}_{n=1}}}

\newcommand{\PSRBFbranches   }{\result{\ensuremath{0.120 \pm 0.002}}} 
\newcommand{\OldBFbranches   }{\result{\ensuremath{0.220 \pm 0.007}}} 
\newcommand{\NewBFbranches   }{\result{\ensuremath{0.146 \pm 0.005}}} 
\newcommand{\NewBFbranchesApprox}{\result{6.9}} 
\newcommand{\NewBFbranchesAlt}{\result{\ensuremath{0.185 \pm 0.006}}} 

\newcommand{\NewBFbranchesGivenPSR}{\result{\ensuremath{1.2}}} 

\newcommand{\PSRLikeRatiobranches}{\result{\ensuremath{1.00}}}
\newcommand{\OldLikeRatiobranches}{\result{\ensuremath{0.97}}}
\newcommand{\NewLikeRatiobranches}{\result{\ensuremath{0.60}}}
\newcommand{\NewLikeRatiobranchesAlt}{\result{\ensuremath{0.94}}} 

\newcommand{\BFcs}{\ensuremath{\mathcal{B}^{c_s^2 > c^2/3}_{c_s^2 \leq c^2/3}}}

\newcommand{\LikeRatiocs}{\ensuremath{\max \mathcal{L}^{c_s^2 > c^2/3}_{c_s^2 \leq c^2/3}}}

\newcommand{\PSRBFcs   }{\result{\ensuremath{10.2 \pm 0.5}}} 
\newcommand{\OldBFcs   }{\result{\ensuremath{2220 \pm 790}}} 
\newcommand{\NewBFcs   }{\result{\ensuremath{1000 \pm 340}}} 
\newcommand{\NewBFcsAlt}{\result{\ensuremath{2450 \pm 1820}}} 

\newcommand{\PSRLikeRatiocs}{\result{\ensuremath{1.0}}}
\newcommand{\OldLikeRatiocs}{\result{\ensuremath{50.8}}}
\newcommand{\NewLikeRatiocs}{\result{\ensuremath{26.7}}}
\newcommand{\NewLikeRatiocsAlt}{\result{\ensuremath{72.7}}}

\newcommand{\NewMtranshigh}{\result{\ensuremath{2}}}
\newcommand{\NewMtranslow}{\result{\ensuremath{1}}}
\newcommand{\Newrhotranshigh}{\result{\ensuremath{4.5}}} 
\newcommand{\Newrhotranslow}{\result{\ensuremath{2.2}}} 

\newcommand{\MillerR}{\externalresult{\ensuremath{13.7^{+2.6}_{-1.5}}}} 
\newcommand{\MillerRwithEOS}{\externalresult{\ensuremath{12.28^{+0.60}_{-0.68}}}} 

\newcommand{\MillerRninety}{\externalresult{\ensuremath{14.30^{+4.33}_{-2.97}}}} 
\newcommand{\RileyR }{\externalresult{\ensuremath{12.4^{+1.3}_{-1.0}}}} 
\newcommand{\RileyRninety }{\externalresult{\ensuremath{12.34^{+1.89}_{-1.67}}}} 

\newcommand{\JCromRXray    }{\result{\ensuremath{13.24^{+2.25}_{-1.93}}}} 
\newcommand{\JCromRAll     }{\result{\ensuremath{12.41^{+0.93}_{-1.16}}}} 

\newcommand{\JCromRImprdata}{\result{\ensuremath{2.09}}} 
\newcommand{\JCromRImprEoS }{\result{\ensuremath{3.12}}} 

\newcommand{\JCromrhoc   }{\result{\ensuremath{10.0^{+3.5}_{-3.6}}}} 
\newcommand{\JCromrhocsat}{\result{\ensuremath{3.57^{+1.3}_{-1.3}}}} 

\newcommand{\newNICER}{J0740+6620}
\newcommand{\oldNICER}{J0030+0451}

\newcommand{\Rtyp}{\ensuremath{R_{1.4}}}
\newcommand{\Rtwo}{\ensuremath{R_{2.0}}}

\newcommand{\rhoc}{\rho_{\rm c}}
\newcommand{\rhonuc}{\rho_{\mathrm{nuc}}}
\newcommand{\csq}{\ensuremath{c^2_{s}}}
\newcommand{\csqmax}{\max \left\{c_s^2/c^2\right\}}

\newcommand{\De}{\ensuremath{\Delta}}
\newcommand{\ep}{\ensuremath{\varepsilon}}

\newcommand{\Msolar}{\ensuremath{\mathrm{M}_\odot}}
\newcommand{\Mt}{M_{\mathrm{t}}}
\newcommand{\rhot}{\rho_{\mathrm{t}}}

\newcommand{\Mmax}{\ensuremath{M_{\rm max}}}

\newcommand{\Rskin}{$R_\mathrm{skin}^{^{208}\mathrm{Pb}}$}

\newcommand{\CIT}{\affiliation{TAPIR, California Institute of Technology, Pasadena, California 91125, USA}}
\newcommand{\CITLab}{\affiliation{LIGO Laboratory, California Institute of Technology, Pasadena, CA 91125, USA}}

\begin{document}
\preprint{N3AS-21-010, INT-PUB-21-014}
\title{Impact of the PSR \newNICER\, radius constraint on the properties of high-density matter}
\author{Isaac Legred}
\email{ilegred@caltech.edu}\CIT\CITLab

\author{Katerina Chatziioannou}
\email{kchatziioannou@caltech.edu}\CIT\CITLab

\author{Reed Essick}
\email{reed.essick@gmail.com}
\affiliation{Perimeter Institute for Theoretical Physics, 31 Caroline Street North, Waterloo, Ontario, Canada, N2L 2Y5}

\author{Sophia Han 
(\begin{CJK}{UTF8}{}\begin{SChinese}韩 君\end{SChinese}\end{CJK})
}
\email{sjhan@berkeley.edu}
\affiliation{Institute for Nuclear Theory, University of Washington, Seattle, WA~98195, USA}
\affiliation{Department of Physics, University of California, Berkeley, CA~94720, USA}

\author{Philippe Landry}
\email{plandry@fullerton.edu}
\affiliation{Nicholas \& Lee Begovich Center for Gravitational-Wave Physics \& Astronomy, California State University, Fullerton, 800 N State College Blvd, Fullerton, CA 92831}

\date{\today}

\begin{abstract}
    X-ray pulse profile modeling of PSR J0740+6620, the most massive known pulsar, with data from the NICER and XMM-Newton observatories recently led to a measurement of its radius.
    We investigate this measurement's implications for the neutron star equation of state (EoS), employing a nonparametric EoS model based on Gaussian processes and combining information from other x-ray, radio and gravitational-wave observations of neutron stars. 
    Our analysis mildly disfavors EoSs that support a disconnected hybrid star branch in the mass-radius relation, a proxy for strong phase transitions, with a Bayes factor of \NewBFbranchesApprox. 
    For such EoSs, the transition mass from the hadronic to the hybrid branch is constrained to lie outside  ($\NewMtranslow,\NewMtranshigh$) \Msolar. We also find that the conformal sound-speed bound is violated inside neutron star cores, which implies that the core matter is strongly interacting. The squared sound speed reaches a maximum of $\NewMaxCs\, c^2$ at $\NewRhoMaxCssat$ times nuclear saturation density at 90\% credibility. Since all but the gravitational-wave observations prefer a relatively stiff EoS, PSR J0740+6620's central density is only $\JCromrhocsat$ times nuclear saturation, limiting the density range probed by observations of cold, nonrotating neutron stars in $\beta$-equilibrium.
\end{abstract}
\maketitle

\section{Introduction}
\label{introduction}

The properties and composition of matter at the highest densities achieved in neutron star (NS) cores remain uncertain~\cite{Lattimer:2015nhk,Ozel:2016oaf,Oertel:2016bki,Baym:2017whm}.
The main observational constraints on the equation of state (EoS) of NS matter at densities $\gtrsim 3 \,\rhonuc$, where $\rhonuc=2.8 \times 10^{14} \mathrm{g}/\mathrm{cm}^3$ is the nuclear saturation density, come from radio measurements of the masses of the heaviest known pulsars~\cite{Demorest:2010bx,Fonseca:2016tux,Antoniadis:2013pzd,Cromartie:2019kug,Fonseca:2021wxt}.
These observations place the maximum nonspinning NS mass above $2 \,\Msolar$, which limits the softness of the high-density EoS and tends to decrease the probability of exotic degrees of freedom that reduce the pressure within NS matter. 

Other probes of NS matter are typically less informative about these high densities.
Nuclear calculations and experiments constrain the EoS respectively around~\cite{Drischler:2016djf,Tews:2018kmu,Drischler:2020hwi,Drischler:2020yad,Essick:2020flb} and below~\cite{Fattoyev:2017jql,Reed:2021nqk,Adhikari:2021phr,Essick:2021kjb,Biswas:2021yge} $\rhonuc$.
Recent measurements of the neutron skin thickness of $^{208}$Pb suggest a stiff EoS for densities $\lesssim \rhonuc$~\cite{Reed:2021nqk,Adhikari:2021phr}, though uncertainties are still large and there is potential tension with other laboratory probes~\cite{Roca-Maza:2015eza,Reed:2021nqk,Essick:2021kjb}.
Gravitational wave (GW) observations by LIGO~\cite{TheLIGOScientific:2014jea} and Virgo~\cite{TheVirgo:2014hva} provide information about the tidal properties of merging NSs~\cite{Flanagan:2007ix,Hinderer:2009ca,Chatziioannou:2020pqz}, and have thus far set an upper limit on the stiffness at $\sim 2\,\rhonuc$.
However, they are intrinsically less informative for larger NS masses.
Tidal effects are quantified through the dimensionless tidal deformability $\Lambda$, which scales roughly as $(R/m)^{6}$~\cite{Zhao:2018nyf} for a NS of mass $m$ and radius $R$, implying that the most massive---and thus most compact---NSs exhibit inherently weaker tidal interactions. 
As an result, the very nature of some $\sim 2$--$3 \,\Msolar$ compact objects observed with GWs, such as the primary in GW190425~\cite{Abbott:2020uma} and the secondary in GW190814~\cite{Abbott:2020khf}, cannot be determined beyond a reasonable doubt~\cite{Tan:2020ics,Essick:2020ghc,Dexheimer:2020rlp,Tews:2020ylw,Fattoyev:2020cws,Biswas:2020xna}.
In the same density regime as the GWs, the electromagnetic counterpart to GW170817 may bound the EoS stiffness from below~\cite{Radice:2017lry,Coughlin:2018miv,Radice:2018ozg,Coughlin:2018fis}, though it is subject to significant systematic modeling uncertainty~\cite{Kiuchi:2019lls,Bauswein:2020aag}. 

Another means of probing dense matter is x-ray emission from hotspots on the surface of rotating NSs.
Identifying and modeling modulations in the hotspot lightcurve can be used to measure NS radii.
Initial results obtained by NICER~\cite{Bogdanov:2019ixe,Bogdanov:2019qjb,Bogdanov:2021yip} for PSR \oldNICER~\cite{Miller:2019cac,Riley:2019yda} complement the tidal measurements from GW170817~\cite{TheLIGOScientific:2017qsa,Abbott:2018wiz,Abbott:2018exr}, as they constrain the EoS at $1$-$2\,\rhonuc$~\cite{Lattimer:2000nx}, disfavoring the softest EoSs~\cite{Landry:2020vaw}.
This ensemble of theory, experiment and observation has helped to establish an overall picture of NS matter in the last few years~\cite{Jiang:2019rcw,Raaijmakers:2019dks,Dietrich:2020efo,Landry:2020vaw,Biswas:2020puz}, which is nonetheless still unresolved at high densities.

Recently, a measurement of the radius of the 2.08 \Msolar~pulsar PSR \newNICER~\cite{Cromartie:2019kug,Fonseca:2021wxt} using x-ray data from NICER and XMM-Newton was reported by two independent analyses~\cite{Miller:2021qha,Riley:2021pdl}.
This radius constraint presents a rare glimpse of the properties of the most massive NSs, and a golden opportunity to obtain observational information about the maximum NS mass, $\Mmax$, as well as potential phase transitions in NS cores.
In the context of the preferred hotspot model in each analysis, \cite{Miller:2021qha} finds \MillerR~km and \cite{Riley:2021pdl} obtains \RileyR~km for \newNICER's radius (medians and symmetric 68\% credible intervals).
For context, the inference reported in \cite{Landry:2020vaw} predicts the radius of $2.08\,\Msolar$~NSs to be \externalresult{$12.08^{+0.79}_{-0.98}$} km at the 68\% confidence level.

Observations of the most massive NSs, such as \newNICER, have important implications beyond the EoS.
They inform the NS mass distribution~\cite{OzelPsaltis2012,Kiziltan:2013oja,Alsing:2017bbc,Farrow:2019xnc,FarrChatziioannou2020,Wysocki:2020myz,Chatziioannou:2020msi,Fishbach:2020ryj}, the classification of the heaviest NS candidates observed with GWs~\cite{Essick:2020ghc}, our understanding of the proposed mass gap between NSs and black holes~\cite{Kreidberg:2012ud,Farr:2010tu}, and the characteristics of NS merger remnants~\cite{Bernuzzi:2020tgt} that influence electromagnetic counterpart emission~\cite{Margalit:2017dij,Bauswein:2017vtn}.
The properties of the high-density EoS are also connected to the properties at other density scales through correlations shaped by causality considerations~\cite{Rhoades:1974fn,Kalogera:1996ci}.

To determine the implications of \newNICER's radius measurement for NS matter, we employ a nonparametric model for the NS EoS based on Gaussian processes (GPs), which offers us the flexibility of an analysis that (i) is not tightly linked to specific nuclear models, (ii) can account for phase transitions, including strong first-order phase transitions that result in disconnected stable branches in the mass-radius relation, and (iii) is not subject to the systematic errors that arise with parametrized EoS families described by a finite set of parameters.
Additionally, the nonparametric EoS model allows us to probe a wider range of intra-density correlations in the EoS than parametric models, something especially relevant for the current data set, which targets a wide range of NS densities~\cite{legred-inprep}.

We find that the new \newNICER\, observation pushes the inferred radii and maximum mass for NSs to larger values: we obtain $\Rtyp=\NewROnePointFour$ km for the radius of a $1.4\,\Msolar$ NS and $\Mmax=\NewMmax \,\Msolar$ for the maximum nonrotating NS mass (we quote medians and 90\% highest-probability-density credible regions unless otherwise noted).
Despite significant statistical uncertainties, the inferred NS radii are consistent with being equal over a broad mass range, with a radius difference of $\De R \equiv \Rtwo - \Rtyp = \NewDeltaR$ km between $2.0\,\Msolar$ and $1.4\,\Msolar$ NSs.
This conclusion rules out a large reduction in the radius for massive NSs, a feature that is sometimes characteristic of strong phase transitions in the mass regime of typical NSs.
Further, we find that EoSs with at least one disconnected hybrid star branch in their mass-radius relation are disfavored compared to those with a single stable branch by a factor of approximately \NewBFbranchesApprox.
This supports the current consensus that all dense-matter observations can be accommodated by a standard hadronic EoS, although the possibility of a phase transition remains viable; only the strongest first-order phase transitions produce more than one stable sequence of compact stars.
If, on the other hand, the mass-radius relation has multiple stable branches, the heaviest star on the first stable branch is either $\lesssim \NewMtranslow \,\Msolar$ or $\sim \NewMtranshigh \,\Msolar$. 
Our results disfavor a transition mass in the intermediate mass regime, suggesting that either all NSs observed to date contain exotic cores, or virtually all are purely hadronic.

We also find support for a violation of the conjectured conformal bound on the sound speed $c_s$ in NS matter, $\csq \leq c^2/3$~\cite{Kurkela:2009gj,Cherman:2009tw,Bedaque:2014sqa}, where $c$ is the speed of light.
Such a violation indicates that the sound speed does not rise monotonically to the perturbative QCD limit ($\csq \rightarrow c^2/3$) at asymptotically high densities~\cite{Tews:2018kmu} and signals the presence of strongly interacting matter in NS cores~\cite{Landry:2020vaw}.
The stiff high-density EoS required by the massive pulsar observations already put the conformal bound in jeopardy~\cite{Bedaque:2014sqa,Tews:2018kmu,McLerran:2018hbz,Alsing:2017bbc,Reed:2019ezm}, but the softer low-density behavior favored by GWs and the NICER radius measurements help reach a Bayes factor of \NewBFcs~(mean and standard deviation from Monte Carlo uncertainty), securely in favor of a violation.
We infer that $\csq$ reaches a maximum of \NewMaxCs\, at a density of $\NewRhoMaxCsText \times 10^{14}\,\mathrm{g}/\mathrm{cm}^3$ ($\NewRhoMaxCssat\, \rhonuc$) in NS matter.

Our results are comparable to other analyses of the new \newNICER~radius measurement.
Reference~\cite{Miller:2021qha} examined the pressure-density relation, the NS radius, and $\Mmax$ using the same set of observational data as we do but did not comment on the possibility of phase transitions in the EoS.
They adopted three different models for the EoS(including a simple, more restricted implementation of a GP) which each yielded different but overlapping constraints on the EoS.
EoS models informed by chiral effective field theory ($\chi$EFT) at low densities and GW170817's electromagnetic counterpart were considered in Refs.~\cite{Raaijmakers:2021uju,Pang:2021jta}; the latter analysis also disfavors EoSs with strong first-order phase transitions, while the former compared two parametric EoS models, finding some model-dependence in their results.
A hybrid nuclear parameterization and piecewise polytrope EoS model was employed in~\cite{Biswas:2021yge}, which also accounted for the recent PREX-II measurement of the neutron skin of $^{208}$Pb~\cite{Adhikari:2021phr}.
Compared to these studies, our less restrictive treatment of the EoS model broadly results in both qualitative and quantitative agreement. Nonetheless, it allows us more freedom to investigate the consequences of the \newNICER~radius measurement for NS matter microphysics, including phase transitions, the conformal sound-speed bound, and the inferred stiffness of the EoS.

The remainder of the paper describes the details and results of our analysis.
In Sec.~\ref{methodology} we briefly describe the methodology we employ as well as the relevant data sets. 
In Sec.~\ref{results1} we present the results of our inference for macroscopic NS properties.
In Sec.~\ref{results2} we discuss the constraints that can be placed on microscopic EoS properties in terms of the sound speed in NS matter and phase transitions.
We conclude and discuss other studies of \newNICER's EoS implications in the literature in more detail in Sec.~\ref{discussion}.

\section{Equation of state inference}
\label{methodology}

Our analysis methodology closely follows that of~\cite{Landry:2020vaw}; here we briefly summarize the main features and discuss the updated treatment of \newNICER. 

\subsection{Hierarchical inference}
\label{sec:hierarchical inference}

In order to combine information from multiple data sets that include statistical uncertainties, we use hierarchical inference~\cite{Loredo:2004nn}.
The relevant formalism and equations are described in detail in Sec. III B of~\cite{Landry:2020vaw}.
The marginal likelihood of each observation (for example, a GW tidal measurement) for a given EoS model is obtained by marginalizing over the relevant parameters for individual events (in the GW case, the binary masses and tidal parameters) assuming some prior distribution (i.e., population model) for the nuisance parameters (in the GW case, the binary masses).
Similar to~\cite{Landry:2020vaw}, we assume a fixed population for all observations given the relatively low number of observations to date.
This simplification also makes the EoS likelihood independent of selection effects~\cite{Mandel:2018mve}.
However, as the size of each data set increases (for example through the observation of additional GW signals), we will need to simultaneously fit the population in order to avoid biases in the EoS inference~\cite{Agathos:2015uaa,Wysocki:2020myz}. 

In the absence of knowledge of the true compact object mass distribution, we choose a uniform population model that extends beyond the maximum mass of all EoSs we consider.
For a given EoS model, we further assume that all objects with $m\leq\Mmax$ are NSs. 
In other words, we assume that it is the EoS, and not the astrophysical formation mechanism, that limits the maximum NS mass. 
Then, for observations of objects known \textit{a priori} to be NSs (such as \newNICER, but unlike the components of the GW events), the normalization of the mass prior mildly penalizes EoSs that predict a maximum mass larger than all observed NS masses. This is an Occam penalty that favors EoSs that occupy a smaller prior volume and do not predict unobserved data in the form of very massive NSs, all else being equal.
If, instead, we truncated the NS mass distribution below $\Mmax$---e.g.,~because we had knowledge of an astrophysical process that limits the maximum NS mass---all EoSs with $\Mmax$ greater than the largest population mass would be assigned equal marginal likelihood.
However, such a choice would have to be accompanied by an arbitrary choice of the truncation mass, given our lack of prior knowledge about the upper limit of the astrophysical NS mass distribution. 

The distinction between these scenarios is important for any analysis of \newNICER, given its high mass. We employ a uniform mass distribution with a lower limit of $0.5\,\Msolar$; hence, an EoS with $\Mmax=3\,\Msolar$ is disfavored in our inference compared to an EoS with $\Mmax=2.5\,\Msolar$ by a factor of $(3-0.5)/(2.5-0.5)=1.25$. In the results presented in later sections, for example Fig.~\ref{fig:corner}, this contributes to the fact that the tail of our $\Mmax$ posterior is slightly tighter than the prior.
More details and a quantitative assessment of the effect of the mass prior are given in the \hyperref[appendix]{Appendix}. 

\subsection{Nonparametric EoS model}
\label{sec:nonpara}

The procedure outlined above requires a model that describes the NS EoS and can be used to compute all relevant macroscopic NS properties, such as masses, radii, and tidal deformabilities~\cite{Oppenheimer:1939ne,Tolman:1939jz,Hinderer:2007mb}.
Following~\cite{Landry:2020vaw}, we use a nonparametric EoS model constructed through GPs conditioned on existing dense-matter EoS models available in the literature; see~\cite{Landry:2018prl,Essick:2019ldf,Landry:2020vaw} for more details.
While the GP never assumes a specific functional form for the EoS, unlike parametric analyses, it does assume probabilistic knowledge about correlations within the EoS.
Each GP is constructed with different hyperparameters that specify a covariance kernel, which in turn controls the scale and strength of correlations between the sound speed at different pressures.
The specific model employed here is described in detail in~\cite{Essick:2019ldf}.
It is constructed as a mixture of $\sim 150$ individual GPs with a broad set of hyperparameters, allowing us to probe a wide range of EoS models with different intra-density correlations. 

The nuclear models on which the process is conditioned contain EoSs with different degrees of freedom, including purely hadronic, hyperonic, and quark models.
We intentionally condition only loosely on these models, resulting in the process termed \textit{model-agnostic} in~\cite{Essick:2019ldf}.\footnote{We emphasize that model-agnostic does not mean model-independent.}
As a result, our EoS prior contains a large variety of EoS behavior, including phase transitions at different density scales and of different strengths; see for example Fig. 1 of~\cite{Essick:2019ldf}. 
This nonparametric approach offers two further advantages over more traditional parametric models~\cite{Read:2008iy,Lindblom:2010bb,Greif:2018njt}: it avoids (i) systematic errors and (ii) strong (and perhaps opaque) intra-density correlations~\cite{legred-inprep} that arise from restricting the EoS to a specific functional form with finite parameters, which will inevitably not match the correct EoS.

To that end, Ref.~\cite{Miller:2021qha} also employed a GP EoS model, citing the same benefits we point out here.
However,~\cite{Miller:2021qha} used a single GP with a single set of hyperparameters (compared to $\sim 150$ GPs we consider) and chose those hyperparameters to approximate the variability observed within tabulated EoSs from the CompOSE database~\cite{CompOSE}.
Therefore, the GP prior explored in~\cite{Miller:2021qha} is more reminiscent of the \textit{model-informed} prior considered in~\cite{Landry:2018prl, Essick:2019ldf} than the \textit{model-agnostic} prior considered here and in \cite{Landry:2018prl, Essick:2019ldf, Landry:2020vaw, Essick:2020flb, Essick:2021kjb}.
In fact, the hyperparameters used in~\cite{Miller:2021qha} assume less variance (smaller $\sigma$) and stronger correlations between pressures (larger $l$) than any of the allowed hyperparameters within our hyperprior (see~\cite{Essick:2019ldf} for more details).
Our results, therefore, intentionally explore broader ranges of possible EoS behavior and intra-density correlations than~\cite{Miller:2021qha}, particularly at high densities where the GP model in~\cite{Miller:2021qha} forces the sound speed to approach the speed of light \textit{a priori}.
Their more closely tailored GP design may explain why Fig.~10 of~\cite{Miller:2021qha} shows that their GP analysis leads to more stringent EoS constraints than parametric EoS inferences.

\subsection{Data}

The data we use are similar to~\cite{Landry:2020vaw,landry_philippe_2021_4678703}, with the addition of the new constraints on the mass and radius of \newNICER.
Specifically, we make use of different combinations of:
(i) the radio mass measurements for J0348+0432~\cite{Antoniadis:2013pzd} and \newNICER~\cite{Cromartie:2019kug,Fonseca:2021wxt};
(ii) the GW mass and tidal deformability measurements from GW170817~\cite{TheLIGOScientific:2017qsa,Abbott:2018wiz,170817samples} and GW190425~\cite{Abbott:2020uma,190425samples}; and (iii) the x-ray mass and radius constraints from \oldNICER~\cite{Miller:2019cac,Riley:2019yda} and \newNICER~\cite{Miller:2021qha,Riley:2021pdl}.
For \oldNICER, we follow~\cite{Landry:2020vaw} and select the 3-spot model from~\cite{Miller:2019cac,miller_m_c_2019_3473466}, though one can obtain very similar bounds with the \oldNICER\, results from~\cite{Riley:2019yda,riley_thomas_e_2019_3386449} instead (see~\cite{Landry:2020vaw}).
As before, we do not assume that any of the binary components of GW170817 and GW190425 were NSs \textit{a priori}.

One notable difference compared to~\cite{Landry:2020vaw} is that \newNICER\, now appears in the list of both radio and x-ray observations.
As described in~\cite{Miller:2021qha,Riley:2021pdl}, the measured mass of \newNICER\, is still dominated by the radio observations~\cite{Fonseca:2021wxt}.
The most recent mass estimate of \externalresult{$2.08^{+0.07}_{-0.07}$} \Msolar~is slightly lower than the originally reported value of \externalresult{$2.14^{+0.10}_{-0.09}$} \Msolar~\cite{Cromartie:2019kug} (68\% confidence level), making it more consistent with other Galactic NS mass measurements~\cite{FarrChatziioannou2020}.
To avoid double-counting, we include \newNICER\, either through its updated mass estimate in the radio list \textit{or} through its mass and radius estimate in the x-ray list.
The difference gives an estimate of the impact of the radius constraint alone on the NS EoS.

When treating \newNICER\, as either a radio or an x-ray observation, we explicitly account for the normalization of the mass prior in Eqs.~(9) and (11) of~\cite{Landry:2020vaw}, in accordance with our choice of fixed population model.\footnote{Unlike in \cite{Landry:2020vaw}, where we assumed the population of NICER targets ended at masses well below $\Mmax$, we assume the population of NICER targets extends well beyond $\Mmax$ and include the proper normalization for the mass prior for both \oldNICER~and \newNICER.}
For the \newNICER\, x-ray data, we use either the NICER+XMM samples from~\cite{Miller:2021qha,miller_m_c_2021_4670689} or the {\tt ST-U} samples from~\cite{Riley:2021pdl,riley_thomas_e_2021_4697625}.
Both sets of samples already incorporate the updated mass estimate from~\cite{Fonseca:2021wxt}, though~\cite{Miller:2021qha} inflates the uncertainty in this measurement by \externalresult{$\pm 0.02\,\Msolar$} out of concern for systematic uncertainties. 
For our analysis, we choose to revert back to the published result from~\cite{Fonseca:2021wxt} and remove the additional uncertainty of $0.04\,\Msolar$. 
In practice, we use the posterior samples from~\cite{Miller:2021qha,miller_m_c_2021_4670689} but weight each sample by ${\cal{N}}(2.08\,\Msolar,0.07\,\Msolar)/{\cal{N}}(2.08\,\Msolar,0.09\,\Msolar)$, the ratio of the inferred mass estimate from~\cite{Fonseca:2021wxt} to the inflated mass estimate used in~\cite{Miller:2021qha}.
This allows for a more direct comparison between the results of~\cite{Miller:2021qha} and~\cite{Riley:2021pdl}. We find a negligible effect on our results when we repeat our analysis with the increased mass uncertainty. 
Table~\ref{tb:lit} summarizes the mass and radius data we use for \newNICER. 

\begin{table}[h]
    \centering
    {\renewcommand{\arraystretch}{1.4}
    \begin{tabular}{lcc} \hline
        Measurement & $m$ [$\Msolar$] & $R$ [km]  \\
        \hline \hline
        Radio \cite{Fonseca:2021wxt}          & \externalresult{$2.08^{+0.11}_{-0.11}$} & - \\
        x-ray NICER+XMM \cite{Miller:2021qha} & \externalresult{$2.07^{+0.11}_{-0.12}$} & \externalresult{\MillerRninety} \\
        x-ray NICER+XMM \cite{Riley:2021pdl}  & \externalresult{$2.07^{+0.11}_{-0.11}$} & \externalresult{\RileyRninety} \\
        \hline
    \end{tabular}
    }
    \caption{
        Measurements of PSR \newNICER's mass and radius used in our inference. Medians and 90\% highest-probability credible intervals are given. For the Miller et al. \cite{Miller:2021qha} measurement we remove the inflated mass uncertainty and convert to a flat-in-radius prior.
    }
    \label{tb:lit}
\end{table}

\begin{figure*}
    \centering
    \includegraphics[width=0.49\textwidth]{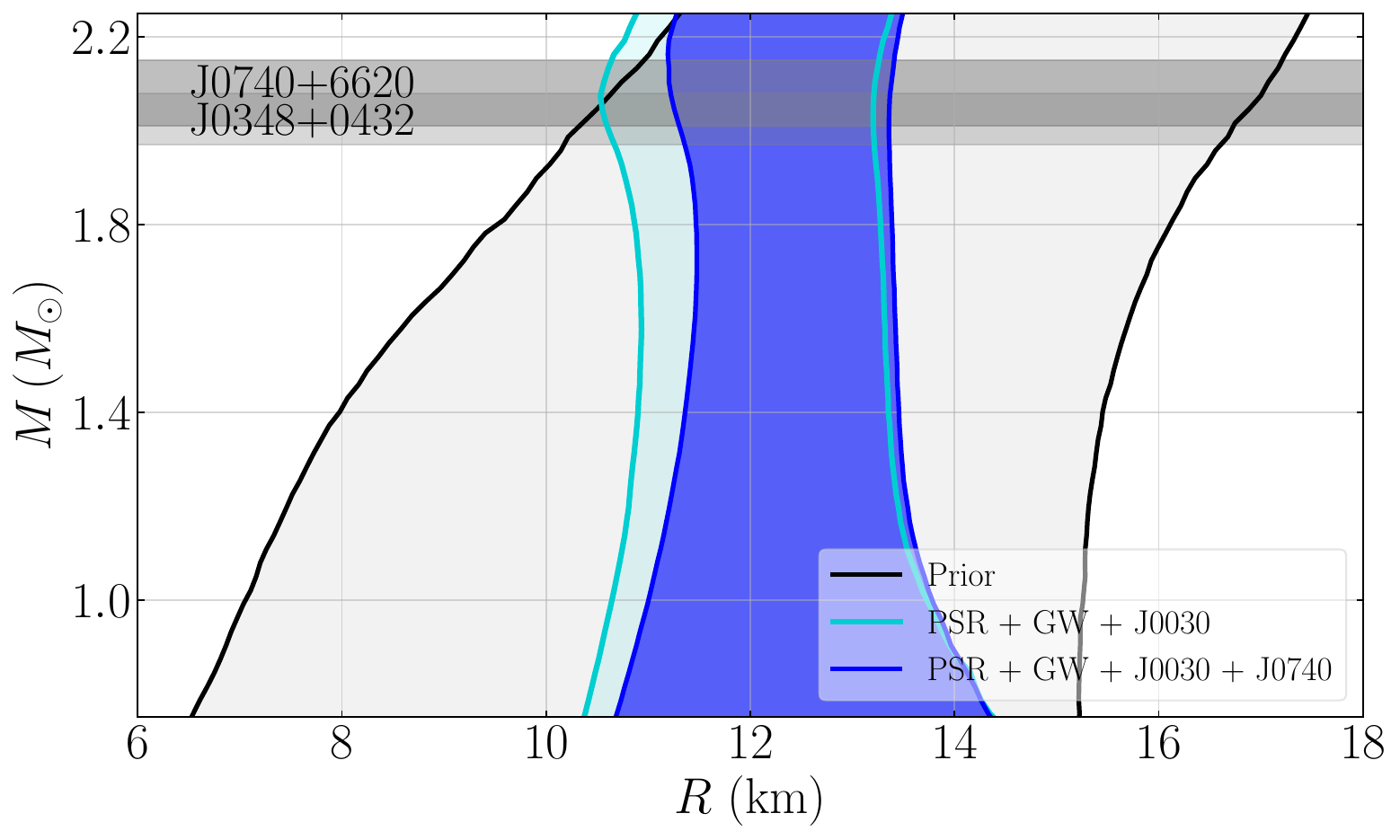}
    \includegraphics[width=0.49\textwidth]{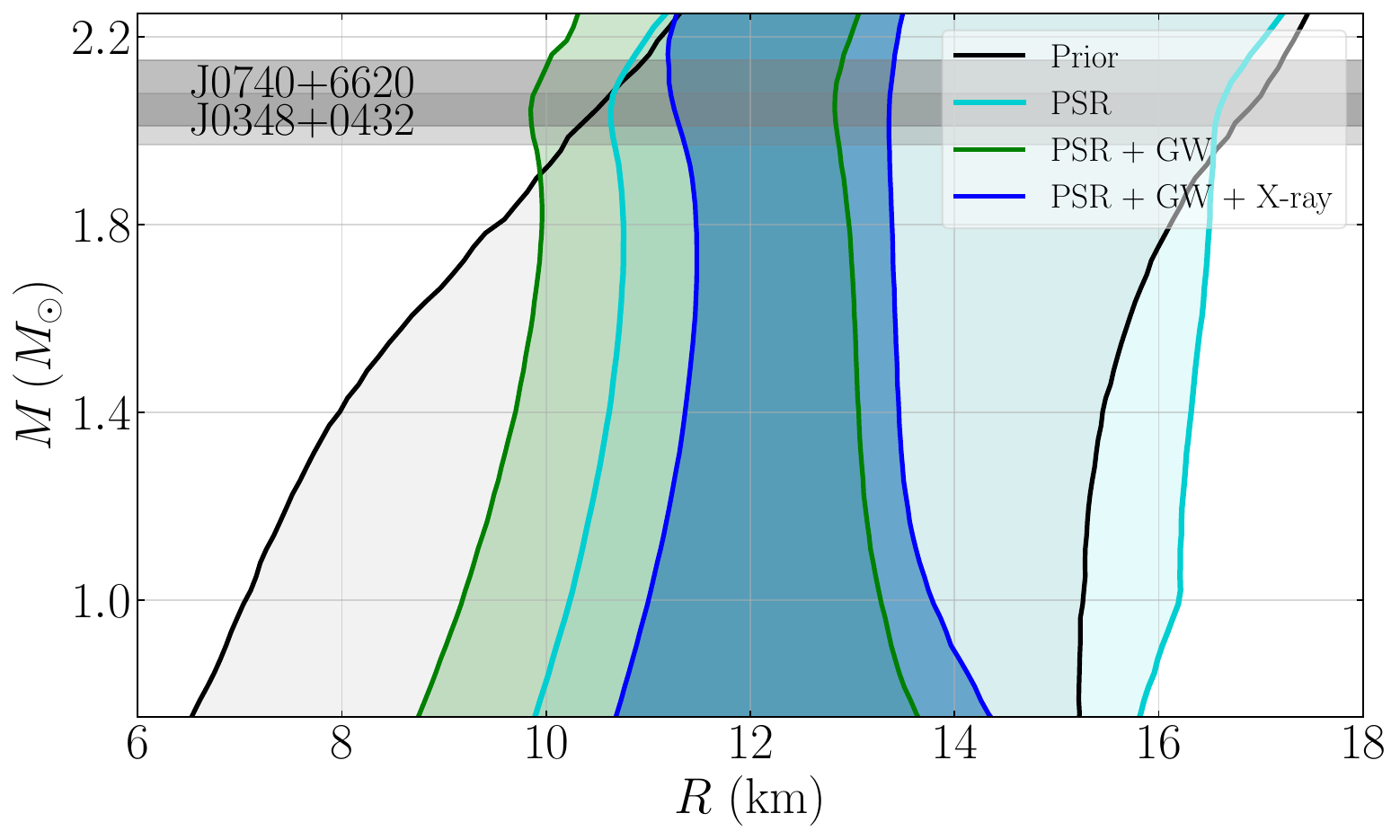}
    \caption{
        Constraints on the NS mass-radius relation.
        Shaded regions enclose the 90\% symmetric credible intervals for the radius for each value of the mass.
        The left panel shows the effect of the \newNICER\, radius constraint by comparing the prior (black), and results with (without) the \newNICER\, radius in blue (turquoise).
        The right panel presents cumulative constraints on the mass-radius relation as each type of data set is analyzed.
        In black we again show the prior.
        The turquoise region shows the posterior after including the mass measurement of the two heavy pulsars (including the updated \newNICER\, mass estimate).
        The green region correspond to constraints obtained after adding the GW data.
        Finally, the blue region correspond to constraints after further adding the \oldNICER\, and \newNICER\, mass and radius constraints from NICER.
        In the last case we remove the \newNICER\, mass constraint from the list of radio constraints so as to avoid double-counting.
    }
    \label{fig:mR}
\end{figure*}

Unlike \oldNICER, the two independent analyses of \newNICER\, arrive at slightly different values for its radius, even if one accounts for their different priors (flat in mass-radius~\cite{Riley:2021pdl} vs. flat in mass-compactness~\cite{Miller:2021qha}) and their different treatments of the uncertainty in the mass estimate from~\cite{Fonseca:2021wxt}.
Accounting for the prior differences \textit{increases} the discrepancy between the two results, as the flat-in-compactness prior disfavors large radii.  
Miller \textit{et al.}~\cite{Miller:2021qha} use the nominal XMM-Newton calibration uncertainty, while Riley \textit{et al.}~\cite{Riley:2021pdl} use a larger uncertainty.
The main effect of the XMM-Newton data is to provide an estimate of the pulsar count rate, which aids in the determination of the relative modulation depth of the x-ray pulse profile, which is essential for placing an upper limit on \newNICER's compactness.
Consequently, the larger calibration uncertainty of~\cite{Riley:2021pdl} results in a less stringent lower bound on the radius.
We focus on results based on the analysis in~\cite{Miller:2021qha}, since it uses the nominal calibration uncertainty, although we provide select comparisons to the results of~\cite{Riley:2021pdl}. 

Nonetheless, we stress that hierarchical EoS constraints are unaffected by the choice of prior for the \newNICER\, radius measurement; any discrepancies are solely due to systematic differences between the two analyses, such as the choice of XMM-Newton calibration uncertainty or issues of convergence within sampling algorithms (see the discussion in Sec.~4.6 of~\cite{Miller:2021qha}).

\section{Neutron star mass and radius}
\label{results1}

We apply our analysis to the combined radio, GW, and x-ray data and present the resulting constraints for macroscopic NS properties, notably masses, radii, and tidal deformabilities.
In what follows, whenever we refer to results without the \newNICER\, radius measurement, we still use its updated mass estimate from~\cite{Fonseca:2021wxt} within the inference.
Unless otherwise stated, all results make use of the Miller et al.~\cite{Miller:2021qha} mass and radius constraints without the inflated mass uncertainty. 

\begin{figure*}
    \centering
    \includegraphics[width=\textwidth]{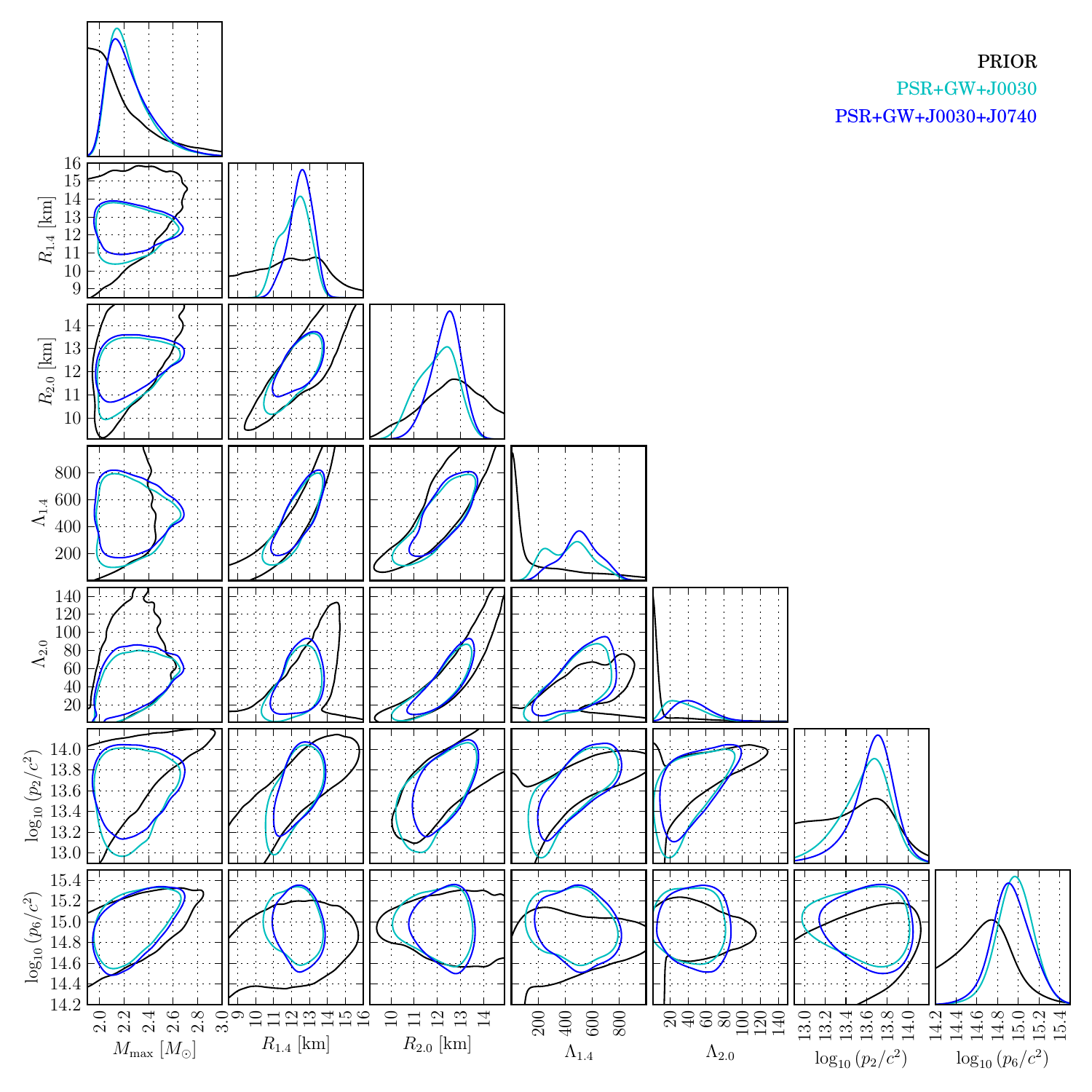}
    \caption{
        Prior and posterior distributions of the radius at $1.4\,\Msolar$ ($\Rtyp$) and $2.0 \,\Msolar$ ($\Rtwo$), the maximum mass ($\Mmax$), the dimensionless tidal deformability at $1.4\,\Msolar$ ($\Lambda_{1.4}$) and $2.0 \,\Msolar$ ($\Lambda_{2.0}$), and the pressure at twice ($p_{2}$) and six times ($p_{6}$) the saturation density,  such that $ p_2/c^2$, and $ p_6/c^2$ have units  $\mathrm{g}/\mathrm{cm}^3$.
        Contours in the 2D distributions correspond to the 90\% level.
        Black lines denote the prior, while blue (turquoise) lines correspond to results with (without) the \newNICER\, radius constraint.
        The prior includes numerous EoSs that do not support massive NSs, in which case we report quantities assuming black holes, corresponding to the sharp peak at $\Lambda=0$ in the prior. 
        }
    \label{fig:corner}
\end{figure*}

We infer the NS mass-radius relation shown in Fig.~\ref{fig:mR}, which plots the 90\% symmetric credible region for $R$ as a function of $m$.\footnote{Fig.~\ref{fig:mR} shows credible regions for $R(m)$ restricted to those EoSs with $\Mmax \geq m$. That is, we show the bounds for stable NSs only.}
The left panel focuses on the effect of the new \newNICER\, radius measurement: it tightens the 90\% credible constraint on the radius from the low side by \result{\ensuremath{\fpeval{round(\NewROnePointFourLow - \OldROnePointFourLow, 2)}}}~km at $1.4\,\Msolar$ and \result{\ensuremath{\fpeval{round(\NewRTwoPointZeroLow - \OldRTwoPointZeroLow, 2)}}}~km at $2.0\,\Msolar$.
The right panel shows cumulative constraints on the mass-radius relation as the different data sets (radio, GW, x-ray) are added one at a time.
As discussed in~\cite{Landry:2020vaw}, the radio and x-ray observations tend to drive the lower bound on the NS radius, while the GW data and causality set the upper bound.
This is because the GW measurements mainly constrain $\Lambda \sim (R/m)^6$ from above, while the x-ray measurements primarily set an upper bound on the compactness $m/R$ and therefore a lower bound on $R$.
The \newNICER\, radius measurement reinforces this picture of complementary constraints.  

One- and two-dimensional marginalized priors and posteriors for various macroscopic and microscopic parameters are given in Fig.~\ref{fig:corner}, while Table~\ref{tab:ResultsSummary} presents medians and 90\% highest-probability-density credible regions for these and other quantities of interest.
Like the left panel of Fig.~\ref{fig:mR}, we compare the prior and posterior with and without the \newNICER\, radius constraint.
However, we no longer restrict the prior to EoSs that support stable NSs at a given mass scale: it includes EoSs with \Mmax~significantly smaller than $1\,\Msolar$, such that, for example, the prior on $\Lambda_{1.4}$ peaks at the black hole value of zero. 
This distinction is less relevant for the posterior, as the data significantly disfavor EoSs that do not support $\Mmax \gtrsim 2\,\Msolar$.
Nonetheless, it explains the shape of some priors in Fig.~\ref{fig:corner}.

On the whole, we find that the \newNICER\, radius constraint increases support for stiffer EoSs with larger radii and tidal deformabilities. Our inferred \Mmax~is also slightly increased.
This is because \newNICER's radius is no smaller than that of a lower-mass NS, indicating that the turning point in the mass-radius relation occurs above the pulsar's mass. As discussed above, the tail of the $\Mmax$ posterior is slightly lower than 
its prior. This is driven by two factors: first the bound on $\Rtyp$ provided by GW170817 that limits $\Mmax$ via causality considerations, and second, our assumption that the maximum NS possible is determined by the EoS and not NS formation mechanisms, 
resulting in EoSs that predict very heavy (and unobserved) NSs being disfavored.
An upper limit on $\Mmax\lesssim 2.2-2.6\,\Msolar$ has been proposed by assuming that the electromagnetic counterpart to GW170817
suggests that the merger remnant collapsed to a BH shortly after merger~\cite{Margalit:2017dij,Shibata:2017xdx,Ruiz:2017due,Rezzolla:2017aly,Shibata:2019ctb,LIGOScientific:2019eut}. We 
do not employ this upper limit here (nor any other information from the GW170817 counterpart), and thus our inferred $\Mmax$ extends to higher values. Indeed the data sets we use can only stringently constrain $\Mmax$ from below. The effect of folding in such an upper limit is demonstrated in~\cite{Pang:2021jta}.

Based on Fig.~\ref{fig:corner}, we also see that $\Rtwo$ is more strongly correlated with the pressure at $2 \,\rhonuc$ than at $6 \,\rhonuc$~\cite{Lattimer:2000nx,Drischler:2020fvz}.
Additionally, \newNICER's radius measurement from~\cite{Miller:2021qha} eliminates the bimodality in the posterior on $\Lambda_{1.4}$~\cite{Abbott:2018wiz}, now favoring the (initially subdominant) upper mode at $\sim 500$ rather than the dominant one at $\sim 200$.
This suggests that the EoS lies on the stiff side of the constraints established by GW170817 at intermediate densities.
We expand on this and quantify the implications for NS central densities in Sec.~\ref{results2}.

The general trend in favor of stiffer EoSs also increases the lower bound of the 90\% highest-probability-density credible region for $\Lambda_{1.4}$ (respectively, $\Lambda_{2.0}$) from \result{\ensuremath{\OldLOnePointFourLow}} (\result{\ensuremath{\OldLTwoPointZeroLow}}) to \result{\ensuremath{\NewLOnePointFourLow}} (\result{\ensuremath{\NewLTwoPointZeroLow}}). 
Setting the tidal deformability equal to this lower limit, we can obtain a conservative estimate of the signal-to-noise ratio (SNR) required for a GW observation to confidently detect tidal effects, i.e., bound $\Lambda$ away from zero.
The measurement uncertainty in $\Lambda$ was $\sim 700$ at an SNR of $\sim 33$ for GW170817~\cite{Abbott:2018wiz}.
Assuming that this measurement is typical and that uncertainties scale inversely with the SNR~\cite{Wade:2014vqa}, a back-of-the-envelope estimate suggests that tidal effects can be measured to within $ \NewLOnePointFourLow$ ($\NewLTwoPointZeroLow$) for a binary with masses of 1.4 \Msolar~(2.0 \Msolar) with SNR of 44 (770).
The threshold SNR for $\Lambda_{1.4}$ is within reach of current advanced detectors~\cite{Aasi:2020wya}, although the SNR for $\Lambda_{2.0}$ will require next-generation detectors, consistent with the findings of~\cite{Chen:2020fzm}.

\begin{figure}
    \centering
    \includegraphics[width=1.0\columnwidth, clip=True, trim=0.5cm 1.0cm 0.0cm 0.0cm]{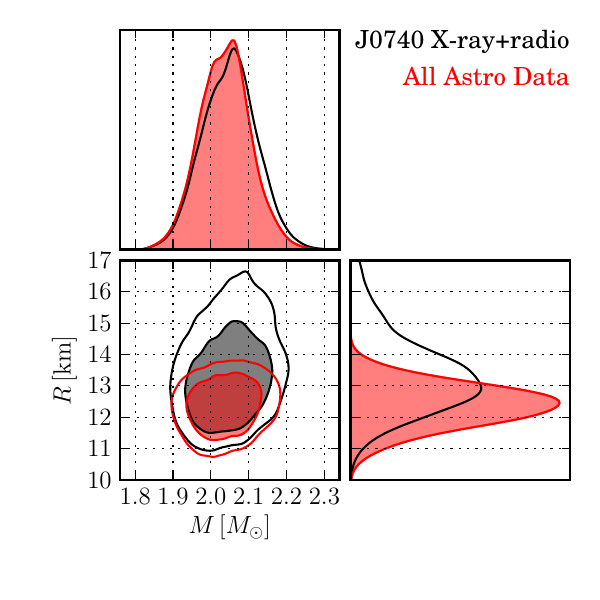}
    \caption{
        Estimates for the radius of \newNICER\, using only NICER+XMM observations (black) and all astrophysical observations (red), both conditioned on our nonparametric EoS representation.
        Contours correspond to the 68\% and 90\% credible levels.
        The primary impact of other astrophysical observations is to lower the inferred radius of \newNICER\, from \JCromRXray~km to \JCromRAll~km at 90\% credibility. \\
    }
    \label{fig:J0740 M-R}
\end{figure}

The full EoS inference also allows us to obtain an updated radius estimate for \newNICER\, informed by all the data, as plotted in Fig.~\ref{fig:J0740 M-R}.
We find \JCromRAll~km at the 90\% level, compared to \JCromRXray~km when using only the \newNICER\, x-ray data conditioned on our nonparametric EoS model.
For reference, the \newNICER~measurement from~\cite{Miller:2021qha} is \MillerRninety~km at 90\% credibility when adjusted to remove the 0.04 \Msolar~systematic error estimate and intrinsic flat-in-compactness prior.
The radius uncertainty for \newNICER\, at the 90\% level is reduced by \JCromRImprEoS~km by conditioning on our EoS prior and further by \JCromRImprdata~km when additionally including all our astrophysical data.
Most of this improvement comes from the exclusion of large radii due to two reasons: (i) the EoS prior model favors realistic EoSs and a radius below $\sim 17$ km, see prior in Fig.~\ref{fig:mR}, and (ii) the GW data are inconsistent with large radii above $\sim 13-14$ km.
The updated radius estimate is consistent with the constraint of \MillerRwithEOS~km from~\cite{Miller:2021qha} (68\% level) after conditioning on other data and their EoS prior.

\begin{table*}[]
    \begin{center}
        {\renewcommand{\arraystretch}{1.4}
        \begin{tabular}{l c @{\quad\quad}c @{\quad}c @{\quad}c @{\quad}c @{\quad}c}
            \hline \hline
             & \multirow{2}{*}{Observable} & \multirow{2}{*}{Prior} & \multirow{2}{*}{w/ PSRs} & \multirow{2}{*}{w/o \newNICER} & \multicolumn{2}{c}{w/\newNICER} \\
            \cline{6-7}
             & & & & & Miller+ & Riley+ \\
            \hline
            & $\Mmax$ $[\Msolar]$    & \PriorMmax    &  \PSRMmax    & \OldMmax      & \NewMmax      & \NewMmaxAlt \\
            \cline{3-7}
            & $p(\rhonuc)$ $[\PnucCoef{\rm dyn}/{\rm cm^2}]$       & \PriorPnuc   & \PSRPnuc  & \OldPnuc      & \NewPnuc          & \NewPnucAlt \\
            \multirow{2}{*}{Properties of} & $p(2\rhonuc)$ $[\PtwonucCoef{\rm dyn}/{\rm cm^2}]$   & \PriorPtwonuc & \PSRPtwonuc  & \OldPtwonuc   & \NewPtwonuc     & \NewPtwonucAlt \\
            \multirow{2}{*}{the EoS} & $p(6\rhonuc)$ $[\PsixnucCoef{\rm dyn}/{\rm cm^2}]$   & \PriorPsixnuc & \PSRPsixnuc & \OldPsixnuc   & \NewPsixnuc      & \NewPsixnucAlt \\
            \cline{3-7}
            & $\max \left\{\csq/c^2\right\} \ \mid \ \rho \leq \rho_c(\Mmax)$ & \PriorMaxCs & \PSRMaxCs & \OldMaxCs & \NewMaxCs  & \NewMaxCsAlt \\
            & $\rho\left(\max \left\{\csq/c^2\right\}\right)$ $[10^{15}\mathrm{g}/\mathrm{cm}^3]$ & \PriorRhoMaxCs & \PSRRhoMaxCs & \OldRhoMaxCs & \NewRhoMaxCs  & \NewRhoMaxCsAlt \\
            & $p\left(\max \left\{\csq/c^2\right\}\right)$ $[10^{35}\mathrm{dyn}/\mathrm{cm}^2]$ & \PriorPMaxCs & \PSRPMaxCs & \OldPMaxCs & \NewPMaxCs  & \NewPMaxCsAlt \\
            \cline{2-7}
            \multirow{2}{*}{Properties defined} & $R_{1.4}$ $[{\rm km}]$ & \PriorROnePointFour & \PSRROnePointFour & \OldROnePointFour & \NewROnePointFour & \NewROnePointFourAlt \\
            \multirow{2}{*}{for both} & $R_{2.0}$ $[{\rm km}]$ & \PriorRTwoPointZero & \PSRRTwoPointZero  & \OldRTwoPointZero & \NewRTwoPointZero & \NewRTwoPointZeroAlt \\
            \multirow{2}{*}{NSs and BHs} & $\De R\equiv R_{2.0}-R_{1.4}$ $[{\rm km}]$ & \PriorDeltaR    & \PSRDeltaR & \OldDeltaR         & \NewDeltaR          & \NewDeltaRAlt \\
            \cline{3-7}
            & $\Lambda_{1.4}$       & \PriorLOnePointFour & \PSRLOnePointFour & \OldLOnePointFour & \NewLOnePointFour            & \NewLOnePointFourAlt \\
            & $\Lambda_{2.0}$       & \PriorLTwoPointZero & \PSRLTwoPointZero & \OldLTwoPointZero & \NewLTwoPointZero            & \NewLTwoPointZeroAlt \\
            \cline{2-7}
            \multirow{2}{*}{Properties defined} & $\rhoc(1.4\,\Msolar)$ $[10^{14}\mathrm{g}/\mathrm{cm}^3]$      & \PriorrhocOnePointFour & \PSRrhocOnePointFour & \OldrhocOnePointFour & \NewrhocOnePointFour            & \NewrhocOnePointFourAlt \\
            \multirow{2}{*}{only for NSs} & $\rhoc(2.0\,\Msolar)$ $[10^{14}\mathrm{g}/\mathrm{cm}^3]$      & \PriorrhocTwoPointZero & \PSRrhocTwoPointZero & \OldrhocTwoPointZero & \NewrhocTwoPointZero            & \NewrhocTwoPointZeroAlt \\
            & $\rhoc(\Mmax)$ $[10^{15}\mathrm{g}/\mathrm{cm}^3]$      & \PriorrhocMmax & \PSRrhocMmax & \OldrhocMmax & \NewrhocMmax            & \NewrhocMmaxAlt \\
            \hline \hline
        \end{tabular}
        }
    \end{center}
    \caption{
        Constraints on selected parameters of interest.
        We present the median and 90\% highest-probability-density credible regions of the marginalized 1D distribution for the maximum mass, the radius, tidal deformability, and central density of a $1.4\,\Msolar$ and a $2.0\,\Msolar$ compact object, the corresponding radius difference, the central density of the maximum-mass NS, the pressure at various densities, the maximum speed of sound, and the pressure and density where the maximum speed of sound is reached.
        For macroscopic observables that are defined for both NSs and BHs, we present credible regions that span both, assuming the Schwarzschild radius ($2GM/c^2$) and $\Lambda=0$ if $m > \Mmax$.
        For properties defined only for NSs, we additionally condition all our distributions on the requirement that $\Mmax \geq m$ so that we only consider EoSs that support stable NSs at $m$.
        Note that this defines slightly different prior distributions for $1.4\,\Msolar$ and $2.0\,\Msolar$ stars, although the point is less relevant for the posteriors.
        The speed of sound is maximized over densities corresponding to stable NSs (below the central density of the \Mmax~stellar configuration: $\rho \leq \rhoc(\Mmax)$), and therefore the exact density range over which we maximize depends on the EoS.
        Columns correspond to the prior, the posterior with only the two heavy pulsars, and the posterior with and without the radius constraint from \newNICER.
        Results with only the two heavy pulsars and without the radius constraint include the updated mass measurement of \newNICER.
        The column with the pulsar-only posterior is similar to the second column of Table IV in~\cite{Landry:2020vaw}. We include it
        as it roughly corresponds to the assumption that all objects up to $\sim2\,\Msolar$ are NSs (as opposed to our prior in some cases).
        We also present results based on both the Miller et al.~\cite{Miller:2021qha} and the Riley et al.~\cite{Riley:2021pdl} analyses.
    }
    \label{tab:ResultsSummary}
\end{table*}

We also investigate how our results change if we use the \newNICER\, data from~\cite{Riley:2021pdl} in place of the data from~\cite{Miller:2021qha}.
The two sets of inferred NS properties are compared in Table~\ref{tab:ResultsSummary}.
Because of their more conservative treatment of calibration error, the Riley et al.~\cite{Riley:2021pdl} data place a less constraining lower bound on \newNICER's radius and therefore result in a more modest shift towards stiff EoSs.
Out of the $\sim 0.8$ km difference between the lower bounds of the 68\% credible intervals on the pulsar's radius obtained by the two analyses, \cite{Miller:2021qha} attributes $0.55$ km to the calibration difference and choices of prior boundaries.
Our hierarchical analysis is immune to the prior difference, and after conditioning on all the observational data we find an overall difference of \result{\ensuremath{\fpeval{round(\NewROnePointFourLow - \NewROnePointFourLowAlt, 2)}}}~km (respectively, \result{\ensuremath{\fpeval{round(\NewRTwoPointZeroLow - \NewRTwoPointZeroLowAlt, 2)}}}~km) in the lower bound of the 90\% credible interval on $\Rtyp$ ($\Rtwo$) due to other systematic differences between~\cite{Miller:2021qha} and~\cite{Riley:2021pdl}.\footnote{The overall difference we find is smaller than the one quoted in~\cite{Miller:2021qha} as we report 90\% and not 68\% levels. The radius distribution for \newNICER\, is fairly asymmetric, so quoting a smaller credible level tends to inflate discrepancies.}

\section{Properties of dense matter}
\label{results2}

\begin{figure*}
    \centering
    \includegraphics[width=0.49\textwidth]{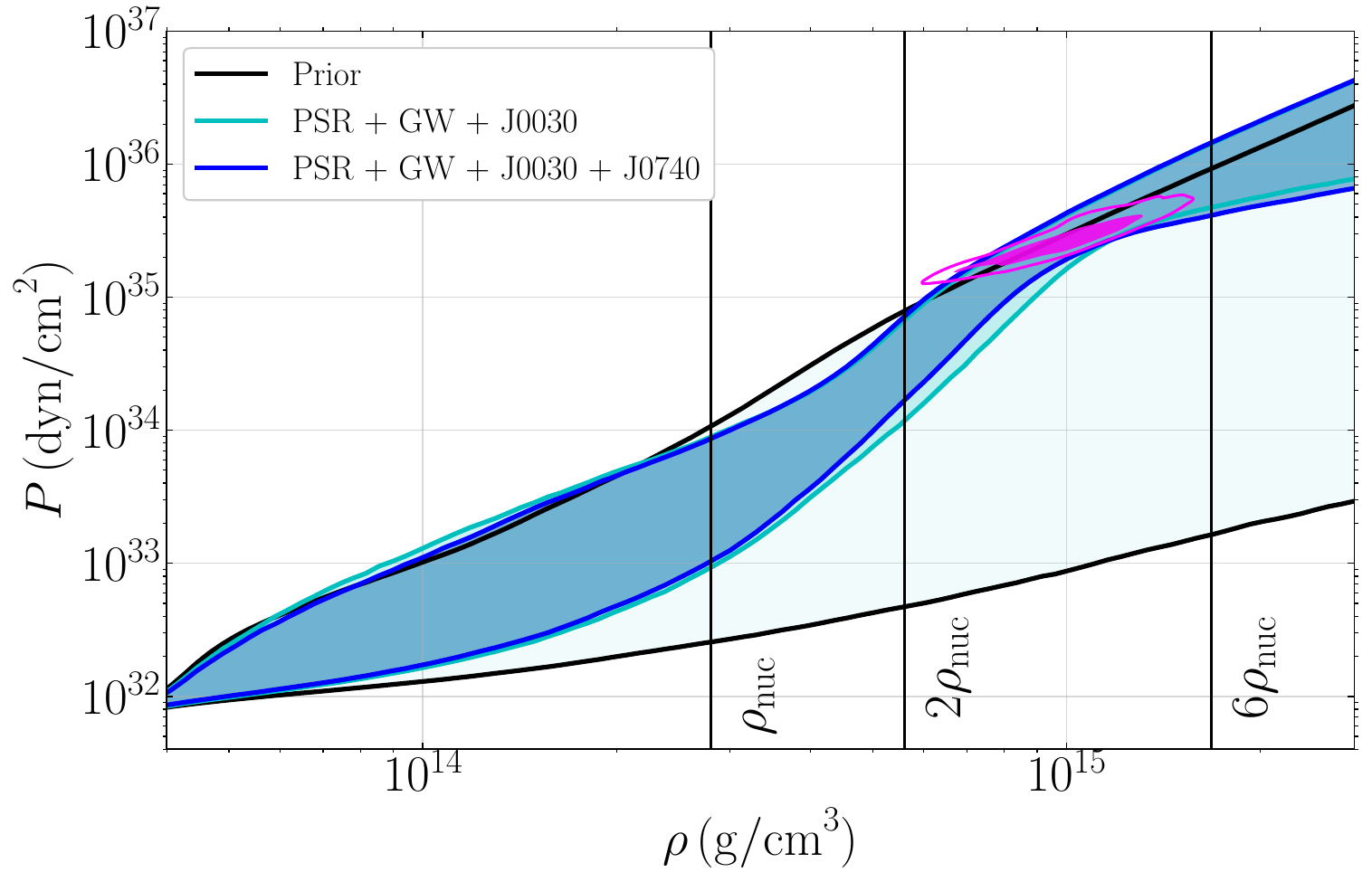}
    \includegraphics[width=0.49\textwidth]{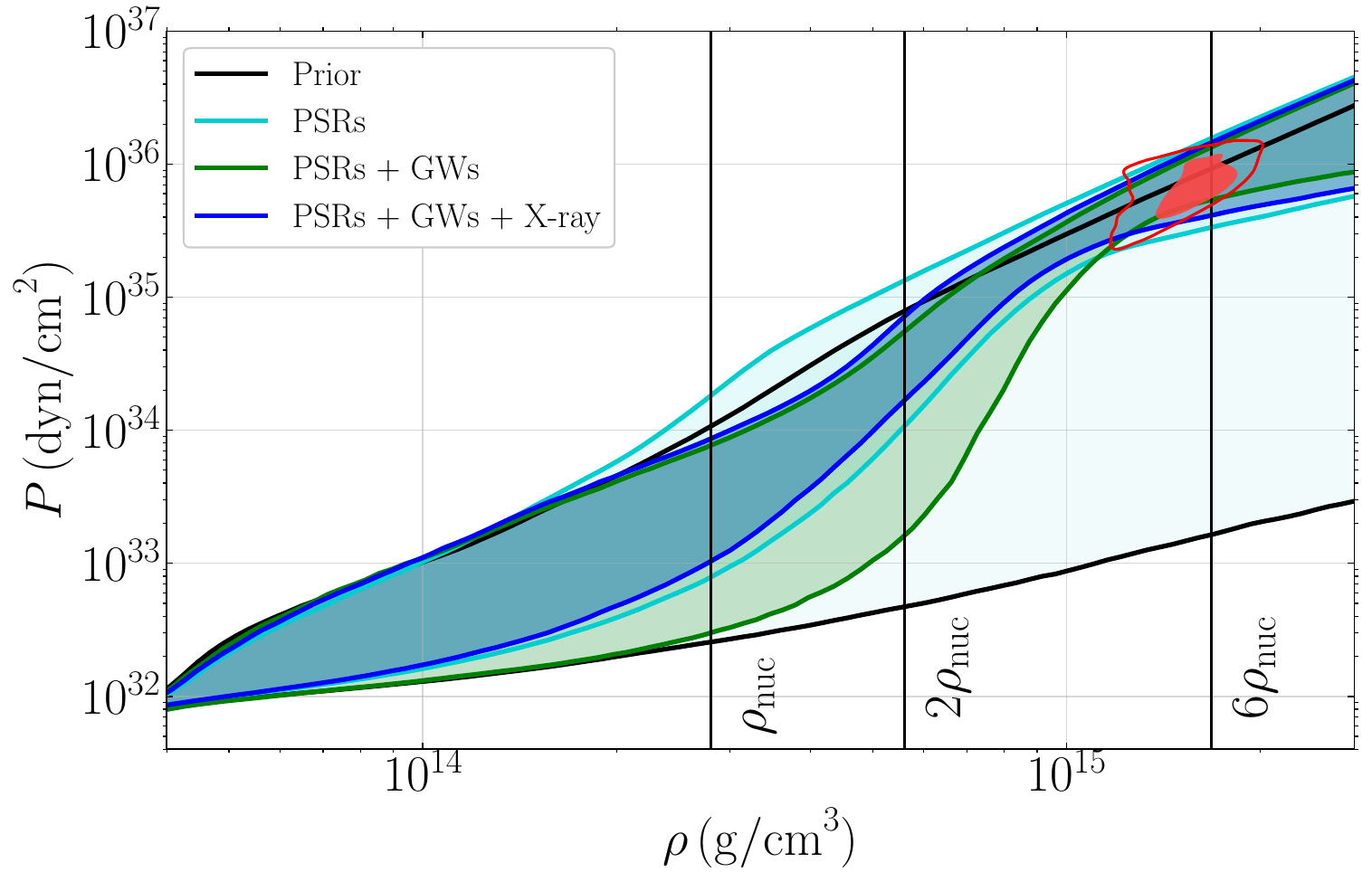} 
    \caption{
        Same as Fig.~\ref{fig:mR} but for the pressure-density relation.
        In the left panel, magenta contours give the 50\% and 90\% level of the central pressure-density posterior for \newNICER\, inferred from all available data. 
        In the right panel, red contours give the 50\% and 90\% level of the central pressure-density posterior for the maximum-mass NS. \\
    }
    \label{fig:prho}
\end{figure*}

We now turn our attention to the properties of dense matter and examine the implications of the \newNICER\, radius constraint. 
We begin in Fig.~\ref{fig:prho} with the inferred pressure-density relation.
In the left panel, we show the effect of the new \newNICER\, radius constraint: it restricts the low-pressure side of the EoS at densities of $2$-$3\,\rhonuc$.
This is comparable, but a bit lower, than the central density of \newNICER, denoted by the magenta contours.
In the right panel, we show the cumulative constraints that result from adding the different data sets sequentially. 
Red contours here denote the central pressure-density posterior for the maximum-mass NS. 

The central density of \newNICER\, is $\JCromrhoc \times 10^{14}\mathrm{g}/\mathrm{cm}^3 \sim \JCromrhocsat \,\rhonuc$, as inferred from all available data under our EoS model.
The relatively low inferred central density for a $\sim 2\,\Msolar$ NS is indicative of a relatively stiff EoS at densities $\sim 1$-$2\,\rhonuc$; see, e.g., Table III of~\cite{Han:2020adu} for a comparison between two representative hadronic models. 
However, our analysis intentionally does not closely follow specific nuclear theoretic predictions.
At low densities (up to $\sim 2\,\rhonuc$), theoretical predictions from $\chi$EFT may place an upper limit on the pressure, which would tend to increase the central density of \newNICER, although the most recent measurement of the neutron skin thickness of $^{208}$Pb~\cite{Adhikari:2021phr} may suggest a relatively stiff EoS below and around $\rhonuc$; see \cite{Essick:2021kjb} for more discussion.\footnote{Fig. 2 of ~\cite{Gandolfi:2011xu} depicts the central densities obtained by extrapolating the realistic two- and three-nucleon interactions predicted by microscopic theory to higher densities. The central values of pressure at around $2\,\rhonuc$ inferred from our analysis (see Table~\ref{tab:ResultsSummary}) point to the stiffest EoS compatible with low-density chiral effective-field-theory ($\chi$EFT)~\cite{Drischler:2020hwi,Drischler:2020yad,Drischler:2020fvz,Raaijmakers:2021uju,Pang:2021jta}.}

\begin{figure*}
    \centering
    \includegraphics[width=0.49\textwidth]{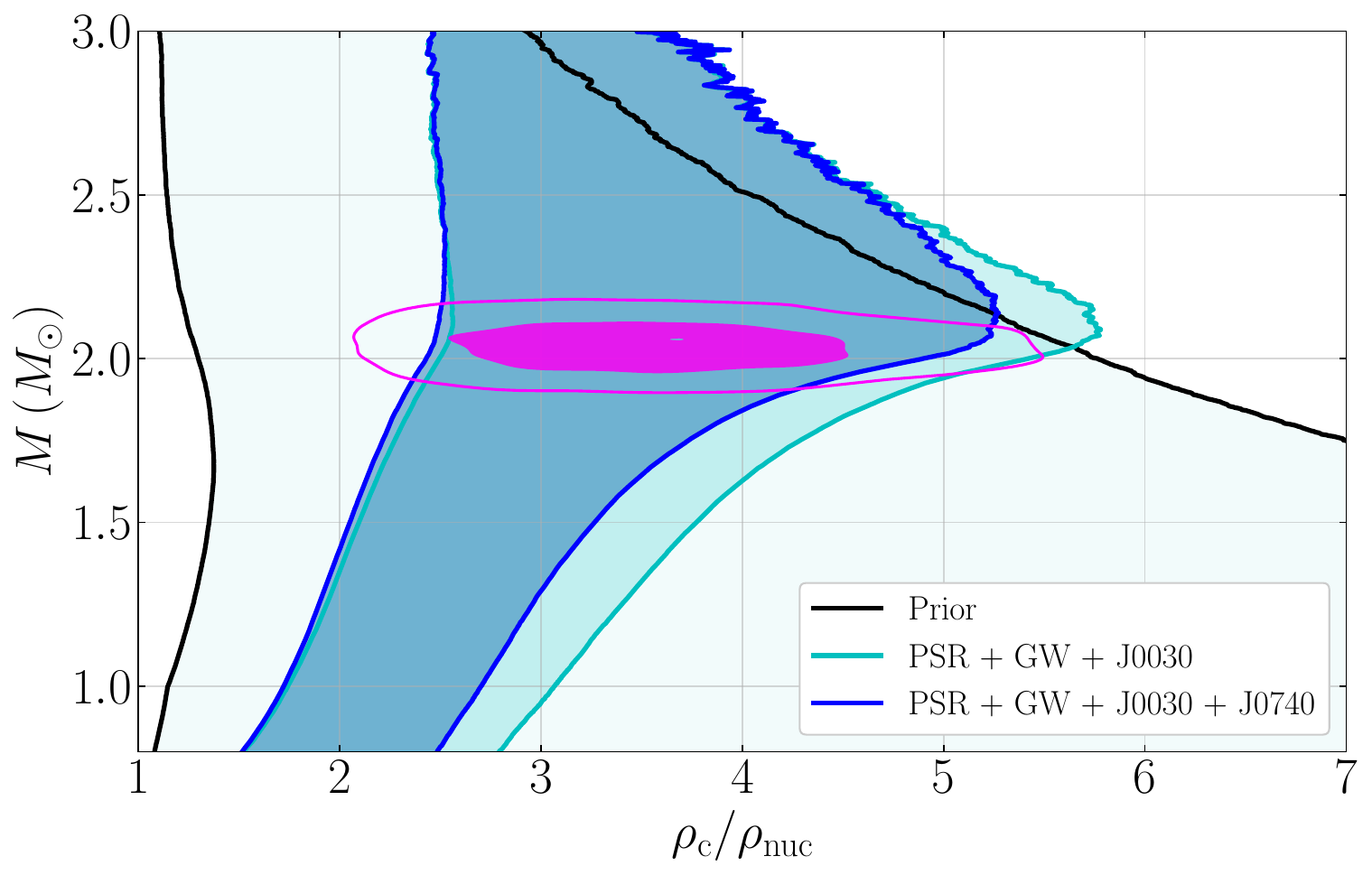}
    \includegraphics[width=0.49\textwidth]{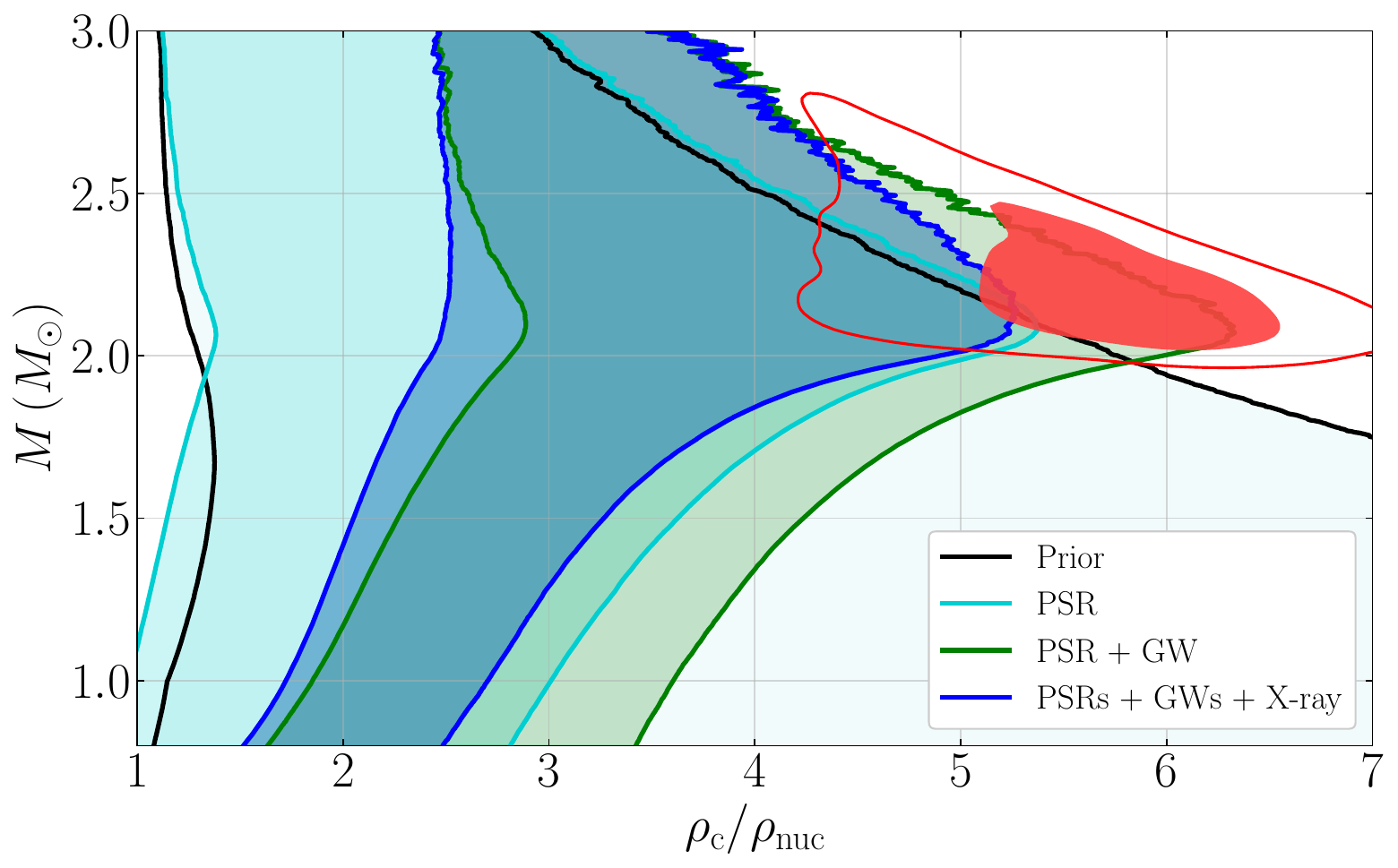}
    \caption{
        Same as Figs.~\ref{fig:mR} and~\ref{fig:prho} but for the central baryon density (in units of saturation density) as a function of NS mass. Magenta (red) contours in the left (right) panel show the 50\% and 90\% credible level for the mass-central density
        posterior for \newNICER\, (maximum-mass NS).
    }
    \label{fig:M-rhoc}
\end{figure*}

We further investigate the NS central densities in Fig.~\ref{fig:M-rhoc}, which shows the mass-central density posterior inferred using all the data.
The central density of the maximum-mass NS is $\NewrhocMmax \times 10^{15}\mathrm{g}/\mathrm{cm}^3 \sim \NewrhocMmaxsat \,\rhonuc$, corresponding to the maximum matter density that can be probed with observations of cold, nonspinning NSs.
Table~\ref{tab:ResultsSummary} also gives the central densities for NSs of $1.4\,\Msolar$ and $2.0\,\Msolar$.
In general, we can understand the trends in the central densities within the same context as Figs.~\ref{fig:mR} and~\ref{fig:prho}.
Typically, the central density remains low (stiff EoS) until masses are $\gtrsim 2\,\Msolar$.
Beyond this limit, set primarily by \newNICER, the EoS can soften appreciably and the central density can increase considerably.
Indeed, the density range explored by NSs above $2\,\Msolar$ could be a factor of two times larger than what is explored by canonical $1.4\,\Msolar$ stars.
High-mass NSs may yet have surprises in store for future measurements.

\subsection{Speed of sound}
\label{sec:speed of sound}

\begin{figure}
    \centering
    \includegraphics[width=1.0\columnwidth]{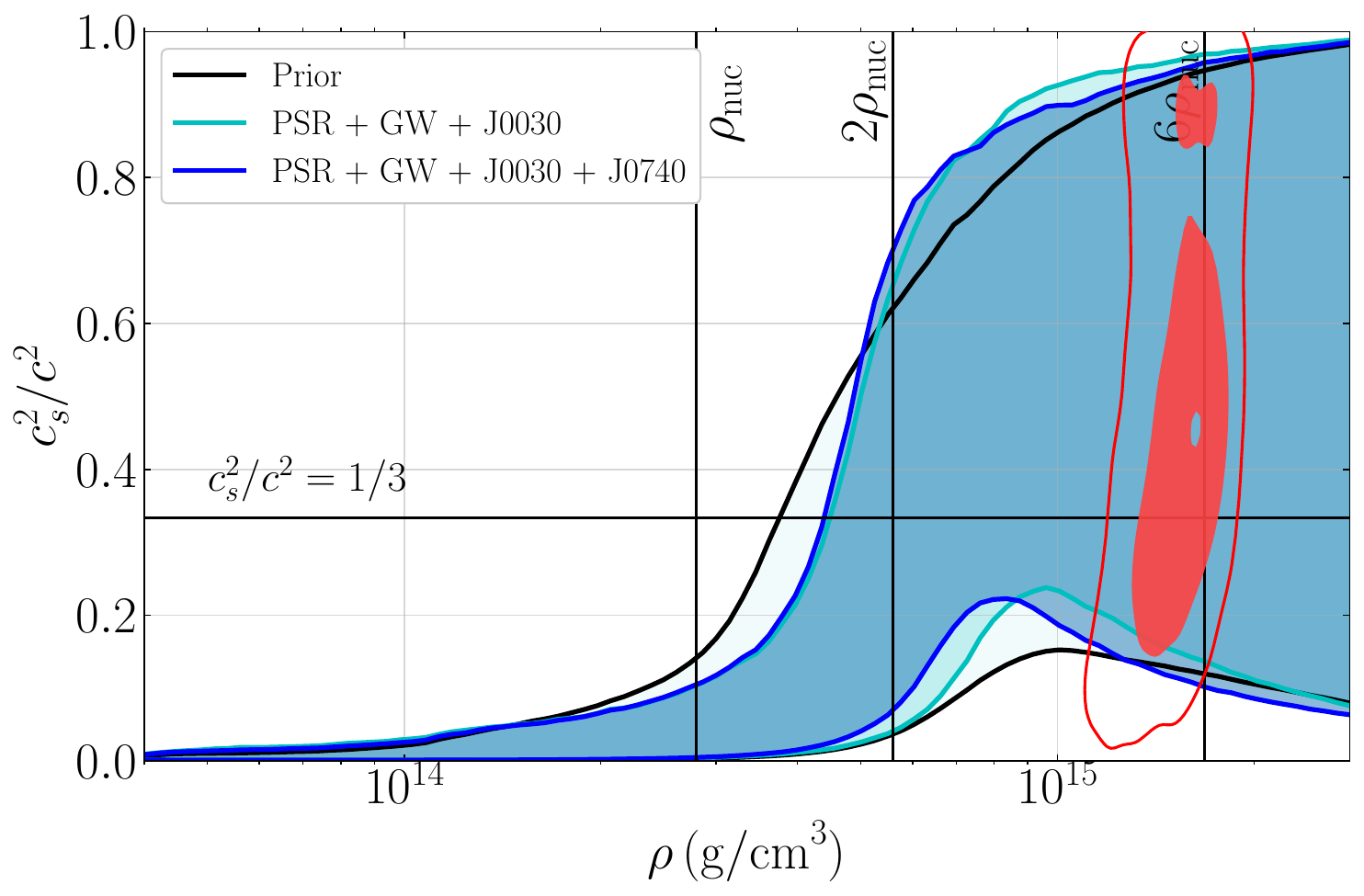}
    \caption{
        Similar to the left panels of Figs.~\ref{fig:mR},~\ref{fig:prho}, and~\ref{fig:M-rhoc} but for the speed of sound inside NSs.
        The horizontal black line denotes the conformal limit $\csq/c^2=1/3$ and the red contour corresponds to the 50\% and 90\% inferred speed of sound and central density for the maximum-mass NS. \\
    }
    \label{fig:sos}
\end{figure}

\begin{figure}
    \centering
    \includegraphics[width=.49\textwidth]{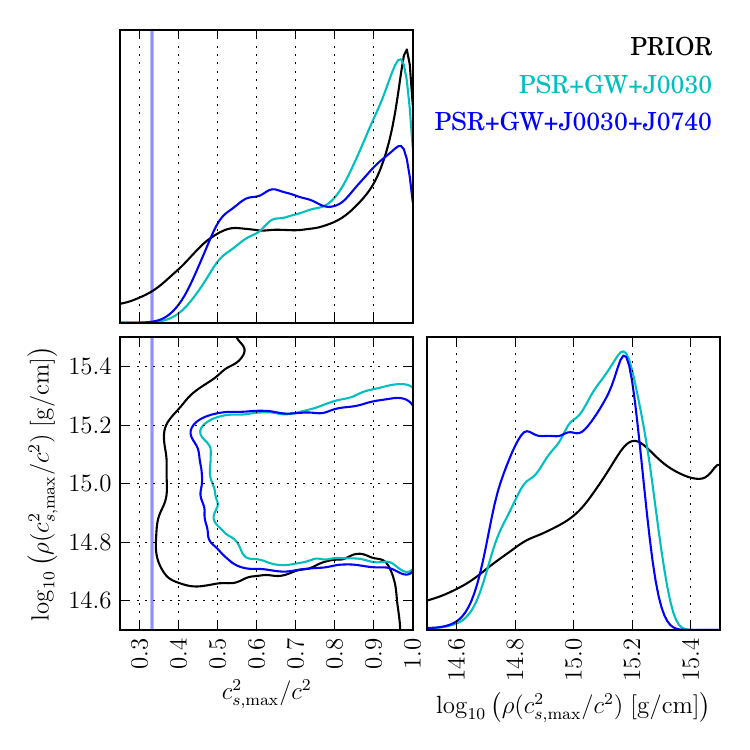}
    \caption{
        One- and two-dimensional marginalized prior and posterior of the maximum $\csq/c^2$ encountered inside the NS and the density at which this happens.
        We show the prior in black and the posterior with (without) the \newNICER\, radius measurement in blue (turquoise).
        The vertical line denotes the conformal limit of $\csq/c^2=1/3$.
    }
    \label{fig:maxsos}
\end{figure}

We examine the speed of sound inside NSs in Figs.~\ref{fig:sos} and~\ref{fig:maxsos}.
Figure~\ref{fig:sos} shows the speed of sound squared ($\csq$) as a function of density with and without the \newNICER\, radius measurement.
Already in~\cite{Landry:2020vaw}, we concluded that the conformal limit of $\csq/c^2=1/3$ is likely violated inside NSs, primarily due to the combination of a soft low-density and a stiff high-density EoS and in agreement with~\cite{Bedaque:2014sqa,Tews:2018kmu,McLerran:2018hbz,Alsing:2017bbc,Reed:2019ezm}.
We here find that the lower limit on the \newNICER\, radius agrees with this picture and pushes the maximum of the marginal 90\% lower limit for $\csq$ to lower densities.
In other words, the pressure needs to increase more rapidly at even lower densities in order to accommodate the relatively large radius of \newNICER. The red contours corresponds to the central speed of sound and central density of the maximum-mass NS, again bounding the densities that can be probed observationally. The central speed of sound is essentially unconstrained, which means that, for some EoSs, the speed of sound sharply decreases after it reaches its maximum value.


Figure~\ref{fig:maxsos} shows the maximum $\csq$ inside NSs and the density at which it is reached.
For each EoS, we maximize $\csq$ over all densities smaller than the central density of the maximum-mass stellar configuration (i.e., a different range for each EoS).
Comparing the posterior to the conformal limit, we again find that the latter is violated inside NSs with a maximum $\csq/c^2$ of \NewMaxCs\, achieved at a density of $\NewRhoMaxCsText \times 10^{14}\mathrm{g}/\mathrm{cm}^3$ ($\NewRhoMaxCssat\,\rhonuc$).
Compared to results without \newNICER, the maximum speed of sound is slightly lower and occurs at slightly lower densities, as also seen in Fig.~\ref{fig:sos}.
This behavior was also observed for \oldNICER~\cite{Landry:2020vaw}. Since
both \oldNICER\, and \newNICER\, data place a lower limit on the NS radius we interpret the reduced value for the maximum speed of sound 
as follows: the preference for a stiffer EoS at $\sim 2\rhonuc$ means that the stiff EoS at $\sim 5 \rhonuc$ can be achieved 
with a milder pressure-density slope and thus a smaller speed of sound. The strongest support for a large speed of sound comes
from the combination of GW and heavy pulsar data that point to a soft low-density and stiff high-density EoS respectively,  
thus necessitating a steep slope in 
between.
Figure~\ref{fig:maxsos} also shows our prior on the maximum $\csq$ and, even though it is consistent with the conformal limit, it 
certainly disfavors it.

To further assess the impact of data on the conformal limit in relation to the prior, Table~\ref{tab:bayes factors} compares the evidence for EoSs that violate the conformal limit ($\max \csq > c^2/3$) with those that obey the conformal limit within nonrotating NSs through the corresponding Bayes factor:
\begin{equation}\label{eq:BFcs}
    \BFcs \equiv \frac{p(\mathrm{data}|\max \csq > c^2/3)}{p(\mathrm{data}|\max \csq \leq c^2/3)} \ .
\end{equation}
We find strong support that the conformal limit is violated: $\BFcs \gtrsim 10^3$.
Although our prior is consistent with EoSs that obey the conformal limit, it includes relatively few realizations that do so.
As such, our Bayes factors are subject to sizeable sampling uncertainty from the finite number of Monte Carlo samples we employ, 
making it hard to conclude whether support for the violation of the conformal limit increases or decreases due to \newNICER.
Nonetheless, we recover large Bayes factors, even considering this sampling uncertainty, and typically find that $\BFcs > 1$ at the $3\sigma$ level.

Similarly, we report the ratio of the maximum likelihood observed for each type of EoS
\begin{equation}\label{eq:DeltaLogLcs}
    \LikeRatiocs = \frac{\max\limits_{\max \csq > c^2/3} \ p(\mathrm{data}|\mathrm{EoS})}{\max\limits_{\max \csq \leq c^2/3} \ p(\mathrm{data}|\mathrm{EoS})}
\end{equation}
This measures how well each type of EoS is able to fit the observed data, and Table~\ref{tab:bayes factors} shows that EoSs that violate the conformal limit are typically favored over those that obey it by between a factor of \result{40--110}.


\begin{table*}
    \centering
    {\renewcommand{\arraystretch}{1.4}
    \begin{tabular}{l l @{\quad\quad}c @{\quad\quad}c | @{\quad\quad}c @{\quad\quad}c}
        \hline \hline
         \multicolumn{2}{c}{Data} & \LikeRatiobranches & \BFbranches & \LikeRatiocs & \BFcs \\
        \hline
        \multicolumn{2}{l}{w/PSRs}
            & \PSRLikeRatiobranches & \PSRBFbranches & \PSRLikeRatiocs & \PSRBFcs \\
        \multicolumn{2}{l}{w/o J0740+6620}
            & \OldLikeRatiobranches & \OldBFbranches & \OldLikeRatiocs & \OldBFcs \\
        \cline{2-6}
        \multirow{2}{*}{w/J0740+6620}
          & Miller+ & \NewLikeRatiobranches & \NewBFbranches & \NewLikeRatiocs & \NewBFcs \\
          & Riley+  & \NewLikeRatiobranchesAlt & \NewBFbranchesAlt & \NewLikeRatiocsAlt & \NewBFcsAlt \\
        \hline \hline
    \end{tabular}
    }
    \caption{
        Ratios of the maximum likelihoods and marginal likelihoods (Bayes factors) comparing EoSs for which the sound speed violates the conformal limit vs. those for which it is satisfied [\LikeRatiocs and \BFcs, Eqs.~(\ref{eq:BFcs}) and (\ref{eq:DeltaLogLcs})], and comparing EoSs with multiple stable branches vs. a single stable branch in their mass-radius relation [\BFbranches~and \LikeRatiobranches, Eqs.~(\ref{eq:BFbranches}) and~(\ref{eq:DeltaLogLbranches})].
        We report point estimates and standard deviation from Monte Carlo sampling uncertainty.
        Data sets are labeled in the same way as in Table~\ref{tab:ResultsSummary}.
    }
    \label{tab:bayes factors}
\end{table*}

\subsection{Strong first-order phase transitions}
\label{sec:numbranches}

\begin{figure}
    \centering
    \includegraphics[width=.49\textwidth]{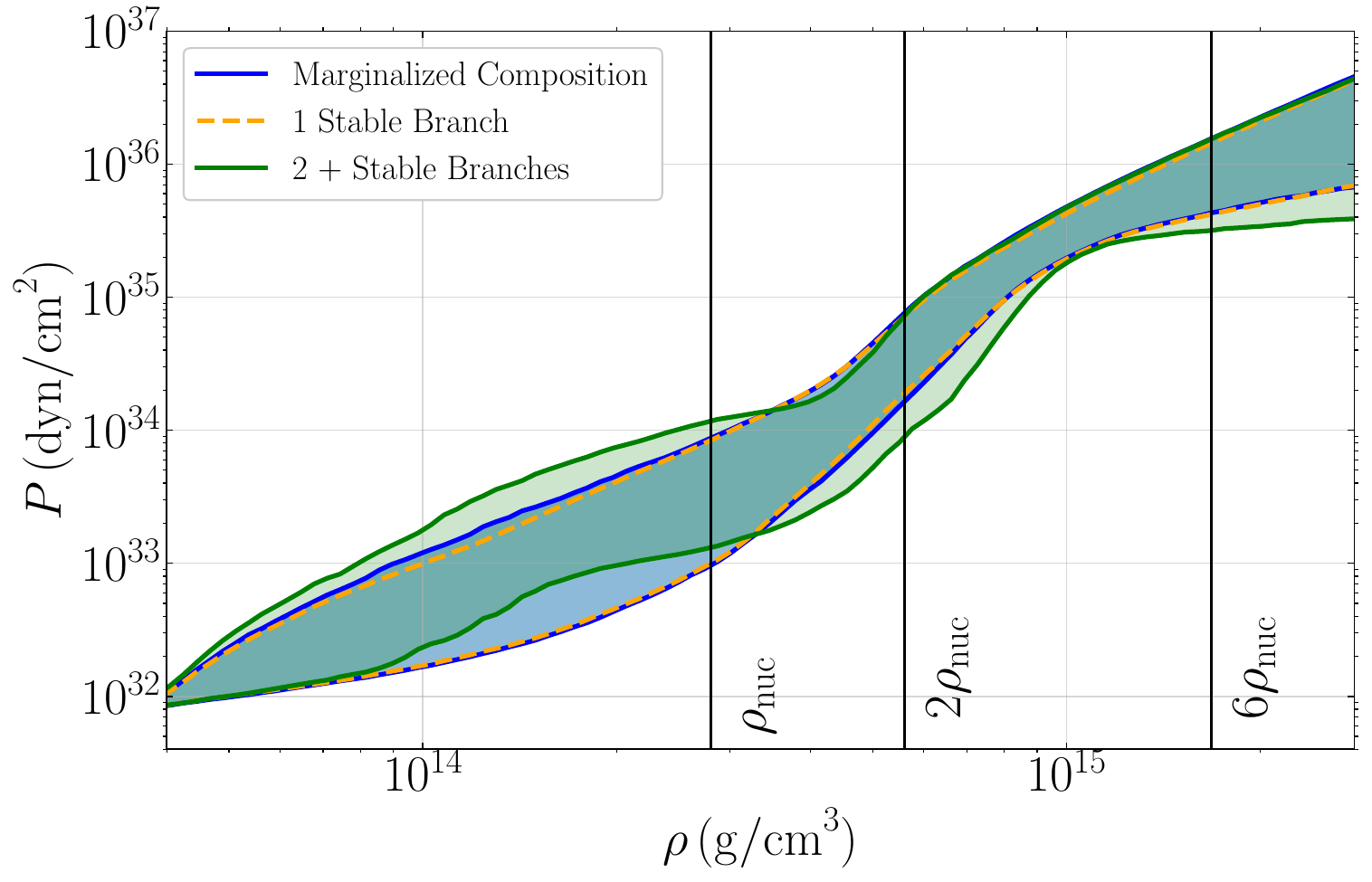}
    \caption{
        Dependence of the pressure-density posterior on the number of stable branches in the EoS.
        The blue region shows the full posterior,
        the green shaded regions show the posterior when restricting to EoSs with multiple stable branches, and the gold dashed lines denote the posterior region when restricting to a single stable branch. 
    }
    \label{fig:prho_branch}
\end{figure}

We now turn our attention to the implications of \newNICER\, for strong phase transitions.
Figure~\ref{fig:prho_branch} compares the pressure-density posterior inferred with EoSs that support different numbers of stable branches
in the mass-radius relation,
used here as a proxy for strong phase transitions.
While strong first-order phase transitions can lead to EoSs with multiple stable branches and possibly even ``twin stars", i.e., stars with roughly the same mass but very different radii~\cite{Schertler:2000xq}, the converse is not necessarily true.
Only the strongest phase transitions lead to disconnected branches, and so what follows concerns only the most extreme phase transitions.
Typically, strong phase transitions and multiple branches result in a large decrease in the radius between subsequent branches~\cite{Alford:2013aca,Alford:2015gna,Han:2018mtj,Chatziioannou:2019yko}.
The lower limit on the radius of \newNICER\, constrains such a sudden decrease. 

Indeed, we find that the full posterior is similar (though not identical) to the one obtained from restricting to EoSs with only a single stable branch.
This suggests that the full posterior marginalized over the number of branches is dominated by EoSs with a single stable branch, though this is also true of the prior. 
Table~\ref{tab:bayes factors} reports the evidence ratios for EoSs with different numbers of stable branches:
\begin{equation}\label{eq:BFbranches}
     \BFbranches \equiv \frac{p(\mathrm{data}|\text{num branches} > 1)}{p(\mathrm{data}|\text{num branches} = 1)} \ .
\end{equation}
We find a Bayes factor of \NewBFbranchesApprox~in favor of a single stable branch, compared to \result{$< 5$} without the \newNICER\, radius measurement.
Astrophysical data generally disfavor the existence of multiple stable branches, driven primarily by the requirement that the EoS supports $\sim 2\,\Msolar$ stars.
As expected, the lower limit on the \newNICER\, further reduces the evidence for multiple stable branches.
However, even the most extreme preference for a single stable branch only suggest a Bayes factor of $\simeq 8$.

Just as with the EoSs that obey vs. violate the conformal limit, we also report ratios in the maximum likelihood observed with EoSs that have a single stable branch vs. those with multiple stable branches
\begin{equation}\label{eq:DeltaLogLbranches}
    \LikeRatiobranches = \frac{\max\limits_{n>1} \ p(\mathrm{data}|\mathrm{EoS})}{\max\limits_{n=1} \ p(\mathrm{data}|\mathrm{EoS})}
\end{equation}
Similar to \BFbranches, we find a preference for EoSs with a single stable branch, but it is small (at most a factor of $\lesssim 2$).

Previous work reported \BFbranches~additionally conditioned on the existence of massive pulsars \textit{a priori}~\cite{Landry:2020vaw}, equivalent to dividing any \BFbranches~by the result using only the massive pulsar observations.
This amounts to examining whether the GW and x-ray data are consistent with multiple stable branches, after we have already assumed the existence of $\sim 2\,\Msolar$ stars.
If we follow suit, we obtain \result{$\BFbranches \sim \NewBFbranchesGivenPSR$} conditioning on the existence of massive PSRs \textit{a priori} and including the x-ray observations of \newNICER, compared to \externalresult{$\sim 1.8$} reported in~\cite{Landry:2020vaw}.
As such, we again find that x-ray observations of \newNICER~lower the evidence in favor of multiple stable branches.
Our conclusions are generally consistent with those reported in~\cite{Pang:2021jta} (\externalresult{$\BFbranches\sim 0.2$}), although our results disfavor multiple stable branches slightly more strongly.
A direct comparison is difficult as~\cite{Pang:2021jta} do not quote uncertainties in their estimates. However, the observed differences could easily be due to priors (e.g., our priors allow for more model freedom and therefore contain more EoSs with multiple stable branches that are not forced \textit{a priori} to support massive stars)
or by how exactly phase transitions are defined (here we define them in terms of stable branches). 

Several caveats should be kept in mind when interpreting our Bayes factors.
Most importantly, it is well documented that Bayes factors are affected by the prior coverage of each model under consideration, particularly if they span regions of parameter space without any support \textit{a posteriori}.
That is to say, the marginal likelihoods that appear in, e.g., Eqs.~(\ref{eq:BFcs}) and~(\ref{eq:BFbranches}) are averages of the likelihood over each prior; if priors span large regions of parameter space with small likelihood values, their marginal evidence will be smaller even if they achieve the same maximum likelihood (match the data just as well) as other, more compact priors.\footnote{For more discussion in a related context, see~\cite{Essick:2020ghc} for a discussion of why posterior odds can be more useful than Bayes factors.}
Indeed, this is why we additionally report \LikeRatiocs~and \LikeRatiobranches.
In particular, differences in the prior support are thought to be a driving factor behind the Bayes factors' apparent preference for EoSs with a single stable branch (multiple-branch EoSs span a broader range of behavior, comparing \LikeRatiobranches~and \BFbranches~in Table~\ref{tab:bayes factors}) as well as the preference for $\chi$EFT models over more agnostic EoS priors~\cite{Essick:2020flb, Essick:2021ezp}.
While this type of Occam factor is desirable in many cases (see, e.g., discussion in Sec.~\ref{sec:hierarchical inference}), one needs to take care when drawing conclusions based on such effects.
Although not guaranteed, we generally find that prior choices of this kind shift our Bayes factor by only a factor of a few, typically much smaller than the variability due to different realizations of experimental noise~\cite{Schad:2021workflow}, which can be as large as factors of ${\cal{O}}(10)$.
It is therefore always prudent to check both the priors and posteriors for the behavior in question, for example checking both Fig.~\ref{fig:maxsos} and Table~\ref{tab:bayes factors} when considering the conformal limit.

In the case of an EoS with multiple stable branches, we find a pressure-density envelope that is morphologically similar to the one in Fig. 4 of~\cite{Landry:2020vaw} (obtained without the \newNICER\, radius data).
These plots show that, if the EoS has multiple stable branches, then the pressure is higher below nuclear saturation and lower at $2$-$3\,\rhonuc$, hinting towards a phase transition in this density regime and suggesting that all observed NSs may already contain an exotic core.
Besides such a low-density phase transition, another possibility is a phase transition at higher densities.
Such an effect is expected to lead to a reduction in the radius of no more than $\sim3$ km~\cite{Han:2018mtj,Chatziioannou:2019yko,Han:2020adu,Drischler:2020fvz} for the most massive NSs compared to \Rtyp, 
which would have been undetectable before the \newNICER\, radius lower limit. 

\begin{figure}
    \centering
    \includegraphics[width=.49\textwidth]{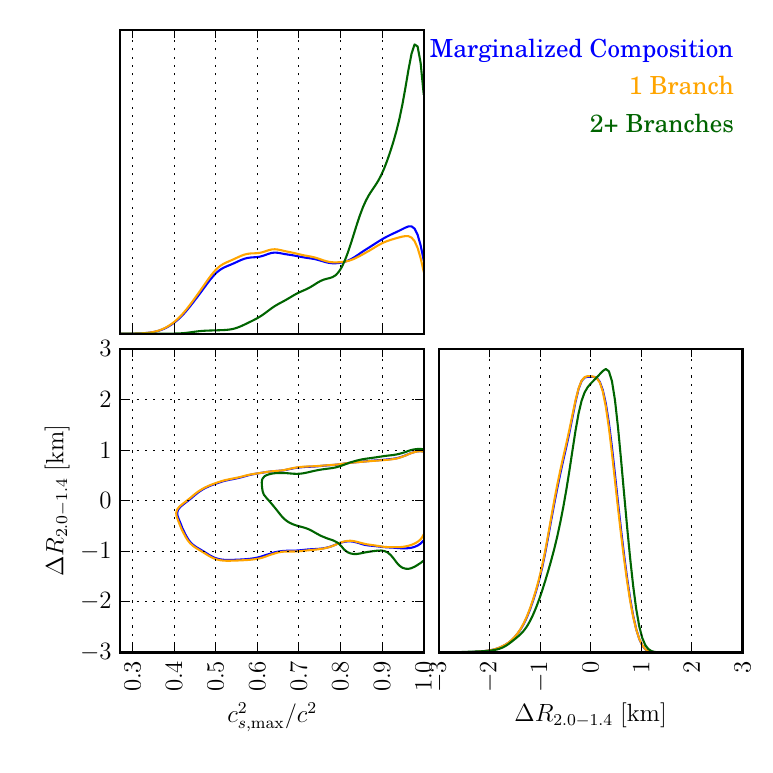}
    \caption{
        Two-dimensional posterior for the radius difference $\De R \equiv \Rtwo-\Rtyp$ against the maximum speed of sound squared reached in the NS.
        We show results with all EoSs (blue) as well as only EoSs with one (gold) and multiple (green) stable branches.
     }
    \label{fig:DeltaR}
\end{figure}

To further explore the implications of a sudden decrease in the radius, in Fig.~\ref{fig:DeltaR} we plot the posterior for the radius difference
$\De R \equiv \Rtwo-\Rtyp$ and the maximum $\csq$, broken down again by the number of stable branches. 
A large negative value for $\De R$ suggests a strong phase transition at low densities, a scenario tightly constrained by the lower limit on the \newNICER\, radius.
We find $\De R=\NewDeltaR$ km, consistent with zero 
although with large uncertainties, as also demonstrated in~\cite{Miller:2021qha,Riley:2021pdl}.
This effectively rules only out the most extreme case of phase transitions that lead to a $\gtrsim 2$ km decrease in radii~\cite{Chatziioannou:2019yko} but still remains consistent with milder or smooth phase transitions~\cite{Han:2020adu,Somasundaram:2021ljr}.
In the case of multiple stable branches, we find that the maximum $\csq$ is higher than the single-branch case, though this does not seem to affect $\De R$.

The larger speed of sound is consistent with previous work that suggests that, in the case of sharp phase transitions, the post-transition speed of sound in general needs to be larger in order to compensate for the intrinsic softening induced by the phase transition~\cite{Han:2018mtj}.
Indeed, as previously studied in~\cite{Drischler:2020fvz}, the absolute bounds on NS radii assuming an EoS with a constant sound speed at high densities is very sensitive to the assumed value of $\csqmax$. The lower (upper) radius bound decreases (increases) as $\csqmax$ increases.
This is in agreement with Fig.~\ref{fig:DeltaR}.
The absolute lower bound on NS radii derived in~\cite{Drischler:2020fvz} corresponds to the most negative value of $\De R$ induced by the strongest possible phase transition (limited by $\csq\leq c^2$ at high densities) compatible with $\Mmax$.
On the other hand, a positive $\De R$ suggests weaker phase transitions, progressing towards the absolute upper bound on NS radii.
We also note that for various physical models of hadronic matter (with or without a smooth crossover to exotic matter), $\De R \gtrsim -1.5$~km is typical, although a few exhibit an increase from $\Rtyp$ to $\Rtwo$~\cite{Han:2019bub,Han:2020adu,Zhao:2020dvu}.

\begin{figure*}
    \centering
   \includegraphics[width=1.0\textwidth]{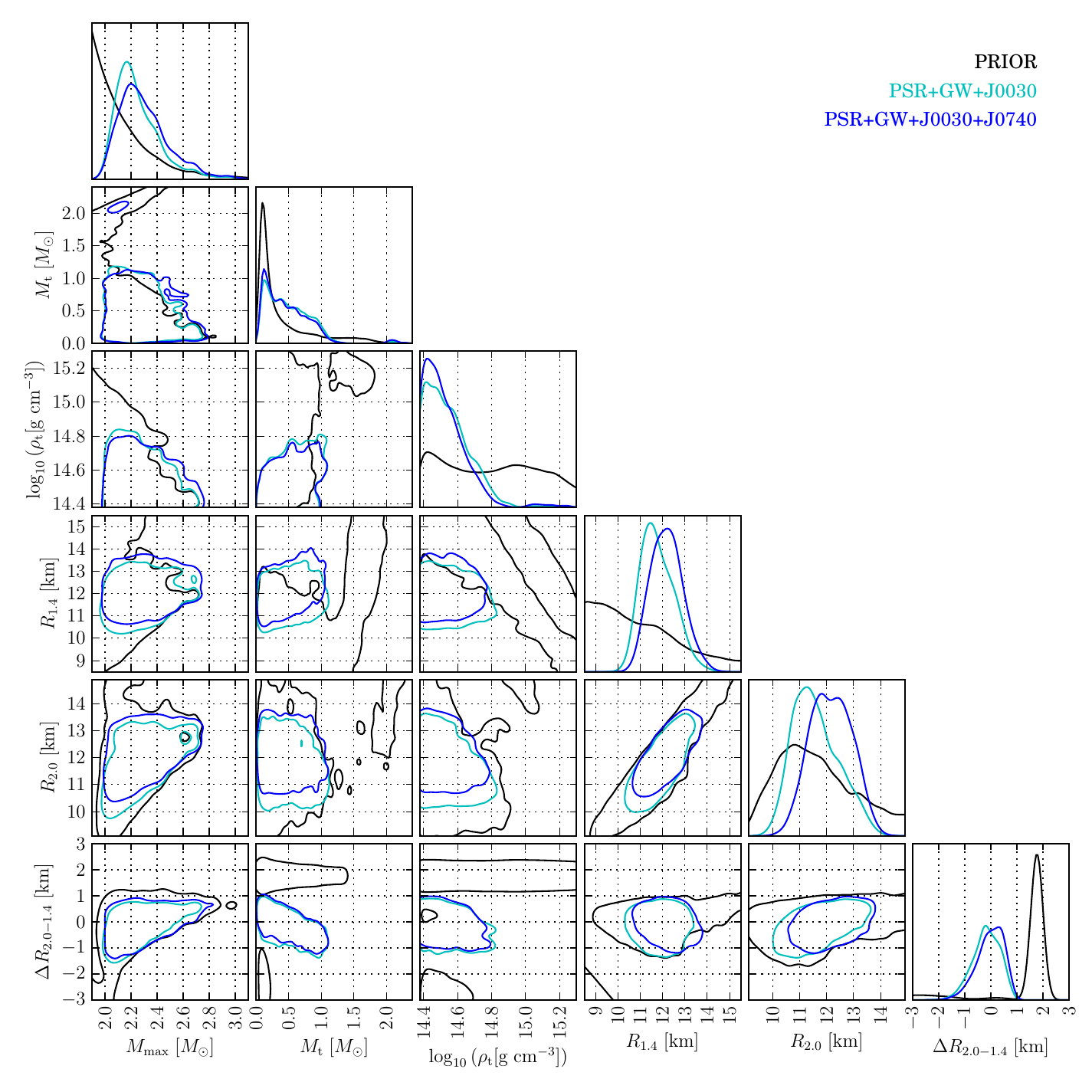}
    \caption{
        Corner plot for the transition mass $\Mt$ corresponding to the most massive hadronic NS in EoSs with multiple stable branches, the transition density $\rhot$ corresponding to the central density of a star with mass $\Mt$, $\Rtyp$, $\Rtwo$, $\Mmax$, and $\De R\equiv \Rtwo-\Rtyp$.
        Contours in the 2D distributions denote the 90\% credible level. 
        Black lines denote the prior, while blue (turquoise) lines correspond to results with (without) the \newNICER\, radius constraint. 
    }
    \label{fig:Mt}
\end{figure*}

To further explore this, in Fig.~\ref{fig:Mt} we plot the posterior for the transition mass $\Mt$, defined as the largest mass of the first stable EoS branch, and the transition density $\rhot$, defined as the central density at the transition mass, 
and select macroscopic quantities. 
Fig.~\ref{fig:Mt} also considers \textit{only} EoSs with multiple stable branches. 
We plot $\Rtyp$, $\Rtwo$, and $\Mmax$, which roughly represent the main observables from GWs, the two NICER pulsars, and the radio mass observations.
We find that, if the EoS has multiple stable branches, the transition from the first branch probably happens for masses $\lesssim \NewMtranslow\,\Msolar$ or $\sim \NewMtranshigh\,\Msolar$.
The corresponding transition density is $\lesssim \Newrhotranslow \,\rhonuc$ or $\sim \Newrhotranshigh \,\rhonuc$.
High-density phase transitions would be the most challenging to detect, as they could result in small changes in the radius and thus be indistinguishable from EoSs without a phase transition~\cite{Alford:2013aca}. 

The posteriors also indicate that transition mass $\Mt$ and the radius difference $\De R$ are anticorrelated.
If the transition mass is very low, then the entire star is mostly composed of exotic matter.
As expected for quark stars, we find that $\De R$ is closer to zero and can even be positive, i.e., the most massive star is bigger (as expected for self-bound configurations).
This is similar to the behavior of the two brown curves in Fig. 1 of~\cite{Chatziioannou:2019yko}.
As the transition mass increases, $\De R$ becomes more negative.
This is similar to the purple and red curves from Fig. 1 of~\cite{Chatziioannou:2019yko} that result in stars that are hadronic in the outer layers but possess a large quark core.
If future GW detections place further upper limits on $\Rtyp$, then large negative values for $\De R$ will be further constrained, thus pushing $\Mt$ even lower. 

The current data disfavor phase transitions that lead to multiple stable branches occurring in the mass range $\sim (1$-$2)\,\Msolar$, suggesting that the majority of NSs we observe belong in a single branch: if the true EoS has multiple stable branches, either all sub-$2\,\Msolar$ contain exotic material or none do.
The two-dimensional $\Mt$-$\Mmax$ plot reveals that this is due to the requirement that $\Mmax \gtrsim 2\,\Msolar$, which disfavors $\Mt \sim 1.5 \,\Msolar$ \textit{a priori} and ``splits" the $\Mt$ posterior into two modes~\cite{Alford:2015gna}.
This behavior is expected, for example, from Fig. 3 of~\cite{Han:2018mtj} which shows that an \Mmax~measurement constrains the intermediate values of the transition pressure.
We leave extraction of further characteristics of the phase transition (such as the transition strength) and EoSs with phase transitions that do not lead to multiple stable branches to future work~\cite{essick-inprep}. 

\begin{figure*}
    \centering
    \includegraphics[width=\textwidth]{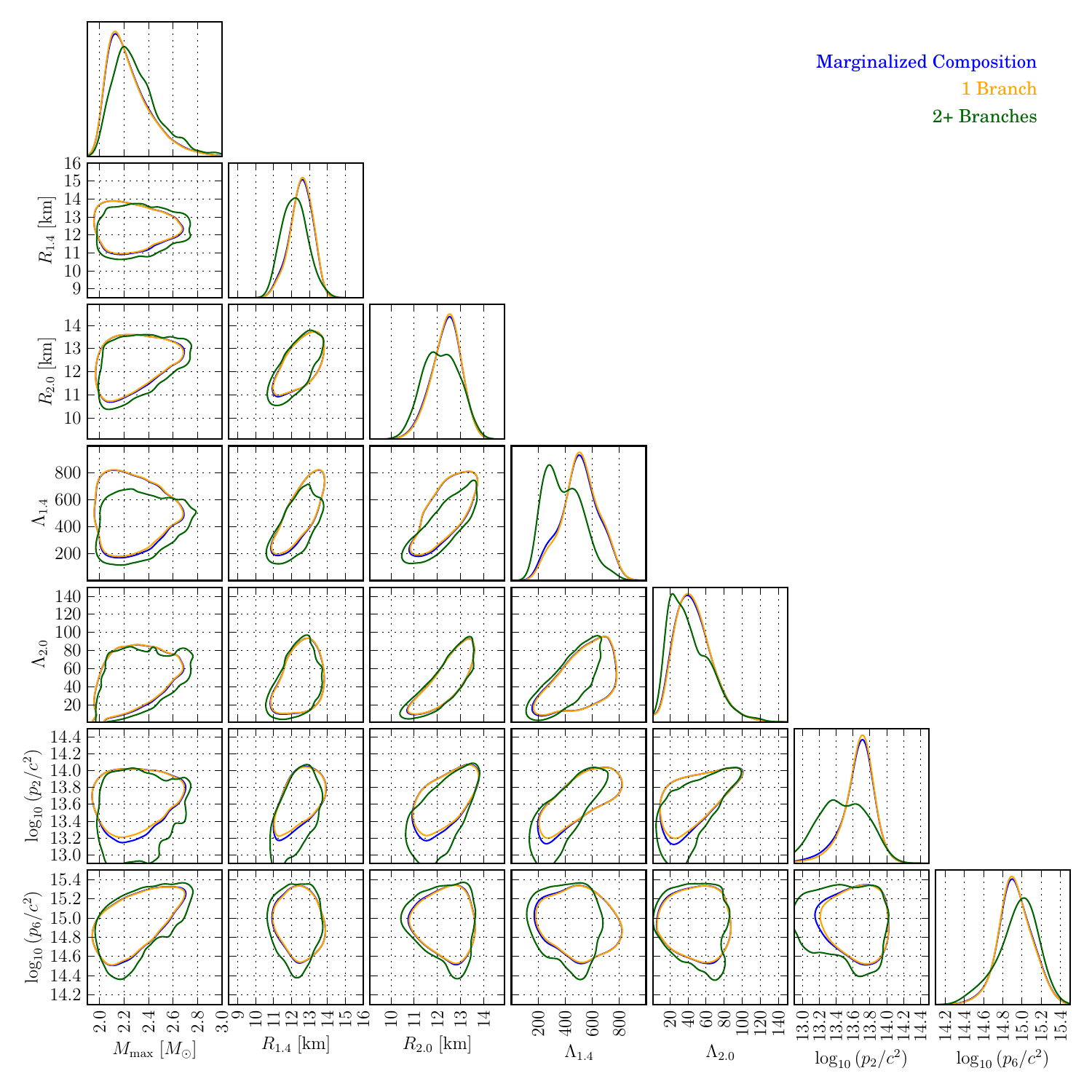}
    \caption{
        Corner plot for various macroscopic and microscopic parameters of interest broken down by number of stable branches.
        Contours in the 2D distributions denote the 90\% level and we plot the same quantities as in Fig.~\ref{fig:corner}.
        Blue lines denote the full posterior, while gold (green) lines correspond to results with EoSs with one (multiple) stable branches. 
    }
    \label{fig:corner_branch}
\end{figure*}

Finally, Fig.~\ref{fig:corner_branch} shows distributions of the same variates as Fig.~\ref{fig:corner}, but separates the EoSs with one and multiple stable branches.
We find that all posteriors are consistent with each other, though EoSs with multiple stable branches are on average consistent with a softer EoS at low densities around $1$-$2\,\rhonuc$ and a stiffer high-density EoS than single-branch EoSs. 
This is expressed through a slightly higher maximum mass and pressure at $6 \,\rhonuc$, but slightly lower radii, tidal parameters, and pressure at $2 \,\rhonuc$. 
The trend towards a stiffer high-density EoS agrees with the maximum speed of sound of Fig.~\ref{fig:DeltaR}.
Similarly, the softer low-density EoS agrees with the pressure-density curves of Fig.~\ref{fig:prho_branch}. \\

Overall, we find that the data mildly disfavor multiple stable branches, though they do not rule out their presence. 
However, if the true EoS indeed has multiple branches, then this would suggest that some extra softening in the EoS, ostensibly due to a phase transition, has already taken place at densities below $\sim 2.2 \,\rhonuc$.
Currently, of all the available astronomical data sets, the GW data dominate the upper limits on the stiffness or pressure of EoS around $2 \,\rhonuc$. 
Should further GW observations continue to push in the same direction, then the evidence for the presence of a strong phase transition in the relevant density region could be strengthened.
To this end, astrophysical observations have limited constraining power at very low densities~\cite{Essick:2020flb, Essick:2021kjb}, and improved theoretical calculations or terrestrial experiments will likely determine whether the pressure is small or large near $\rhonuc$ (see, e.g.,~\cite{Adhikari:2021phr, Roca-Maza:2015eza}).

\section{Discussion}
\label{discussion}

In summary, the new radius measurement for \newNICER\, refines our inference of the EoS by tightening the constraint on the pressure at densities $\sim 2$--$3 \,\rhonuc$.
Like NICER's previous observation of \oldNICER, this constraint comes mainly from the soft side, as the x-ray pulse-profiling available to date primarily bounds NS radii, tides, and pressures from below.
We infer that all observed NSs have the same radius to within $\sim 2$ km.
This picture is consistent with other recent studies of \newNICER~\cite{Miller:2021qha,Raaijmakers:2021uju,Pang:2021jta}.
Our analysis draws three further principal conclusions: (i) the sound speed in NS cores very likely exceeds the conformal bound; (ii) the lack of a large radius difference between high- and low-mass NSs renders the existence of a separate stable branch in the mass-radius relation less likely; and (iii) the stiff EoS around $2 \,\rhonuc$ implied by the ensemble of observations results in a relatively low central density of $\JCromrhocsat \, \rhonuc$ for \newNICER, capping the density range that  astronomical observations of nonrotating NSs can probe to date.
However, the fact that the radius of \newNICER\, is comparable to $\Rtyp$ suggests that \newNICER\, at $\sim2.08\,\Msolar$ might not be at the turning-point of the mass-radius curve and more massive NSs are possible; the inferred central density of the maximum-mass NS is
$\NewrhocMmaxsat \, \rhonuc$.

Our main results are based on the \newNICER\, mass-radius constraint from~\cite{Miller:2021qha}, mainly due to the fact that this analysis uses the nominal relative NICER/XMM-Newton calibration uncertainty.
Nonetheless, we find broadly consistent results when using the data from~\cite{Riley:2021pdl} instead.
The larger calibration uncertainty assumed by~\cite{Riley:2021pdl} results in a weaker lower bound on the radius of \newNICER, and after conditioning on all the observational data this translates to a \result{\ensuremath{\fpeval{round(\NewROnePointFourLow - \NewROnePointFourLowAlt, 2)}}}~km difference in the lower limit of the 90\% credible interval we extract for $\Rtyp$.
Our conclusions about strong phase transitions and the violation of the conformal sound-speed bound are unaltered when the data from~\cite{Riley:2021pdl} is used.
Prior differences in the two analyses (flat-in-compactness~\cite{Miller:2021qha} vs. flat-in-radius~\cite{Riley:2021pdl}) do not affect results within the hierarchical inference formalism.  

A direct numerical comparison between our results and~\cite{Miller:2021qha,Raaijmakers:2021uju,Pang:2021jta,Biswas:2021yge} must be done with care due to the different data sets used and other assumptions.
For example,~\cite{Raaijmakers:2021uju,Pang:2021jta} include GW170817 counterpart models, which we omit here due to concerns about systematic errors, and they assume \textit{a priori} that GW190425 was a binary neutron star merger, which informs the \Mmax~inference because of its large primary mass.
Nonetheless, with those caveats in mind, we can compare posterior constraints on the radius of a 1.4 \Msolar~NS and the maximum NS mass.
Ref.~\cite{Miller:2021qha} finds \externalresult{$\Rtyp = 12.63^{+0.48}_{-0.46}$}\, km and \externalresult{$\Mmax = 2.23^{+0.24}_{-0.15}\,\Msolar$} at the 68\% credible level for their GP model, in very close agreement with our results.
Ref.~\cite{Pang:2021jta} finds \externalresult{$\Rtyp = 12.03^{+0.77}_{-0.87}$}\, km and \externalresult{$\Mmax = 2.18^{+0.15}_{-0.15}\,\Msolar$} at the 90\% credible level, which are smaller and more tightly constrained than our corresponding estimates, using a $\chi$EFT-informed parametric EoS model.
Besides the aforementioned caveats, this difference can be partly attributed to the fact that Ref.~\cite{Pang:2021jta} reports the radius with respect to a flat prior, whereas we report it, like all our constraints, with respect to the prior informed by our nonparametric EoS model.
Both of these results refer to the \newNICER~data from \cite{Miller:2021qha} and can therefore be compared to the second-last column in Table~\ref{tab:ResultsSummary}.
Meanwhile, Ref.~\cite{Raaijmakers:2021uju} reports \externalresult{$\Rtyp = 12.33^{+0.76}_{-0.81}$}\, km and \externalresult{$\Mmax = 2.23^{+0.14}_{-0.23}\,\Msolar$} at the 95\% credible level based on piecewise polytropes informed by $\chi$EFT at low densities, and  Ref.~\cite{Biswas:2021yge} obtains \externalresult{$\Rtyp = 12.61^{+0.36}_{-0.41}$}\, km at the 68\% credible level and a maximum mass of \externalresult{$\sim 2.2\,\Msolar$} using a 
nuclear parameterization for the EoS with a piecewise polytrope extension.
These numbers can be compared to the last column in Table~\ref{tab:ResultsSummary} as they are based on the \newNICER~data from \cite{Riley:2021pdl}.
The results from \cite{Raaijmakers:2021uju} in particular match our inferred values very closely, though our uncertainties are broader, which we attribute to the larger model freedom inherent in our GP prior.
All these results are further broadly consistent with radius estimates from x-ray observations of NSs in low-mass x-ray binaries during quiescent or bursts phases~\cite{Steiner:2017vmg,Nattila:2017wtj,Nattila:2015jra,Ozel:2015fia,Kumar:2019xgp}, though these are subject to considerable modeling uncertainties.

This comparison of our results with the existing literature~\cite{Miller:2021qha,Riley:2021pdl,Raaijmakers:2021uju,Pang:2021jta} brings forward the issue of model dependence in EoS constraints obtained from observations, experiments, and calculations that span many orders of magnitude in density.
By design, the GP EoS prior used in~\cite{Miller:2021qha} does not allow for as much model freedom as our \textit{model-agnostic} process due to the strong intra-density correlations it assumes \textit{a priori}.
This is especially true at high densities.
Another approach to nonparametric inference is to use neural networks, as in~\cite{Han:2021kjx}, though the model constructed in that study deliberately seeks to closely reproduce the behavior of a handful of tabulated EoSs from the literature.
In this sense, the nonparametric models used in~\cite{Han:2021kjx} and~\cite{Miller:2021qha} are more analogous to the \textit{model-informed} GP prior from \cite{Essick:2019ldf} that makes relatively strong prior assumptions about correlations within the EoS.
Parametric EoS models, such as piecewise polytropes~\cite{Read:2008iy}, the spectral decomposition~\cite{Lindblom:2010bb}, and the speed-of-sound parameterization~\cite{Greif:2018njt,Tews:2018kmu}, impose even more restrictive assumptions on EoS morphology by virtue of specifying the functional form of the EoS with a finite number of parameters to describe an infinite-dimensional function space. 
Examples of such model dependence are given in Fig. 10 of~\cite{Miller:2021qha} and the variation between the two models presented in~\cite{Raaijmakers:2021uju}.

These considerations pose the problem of the degree to which EoS constraints are driven by the data, rather than by correlations between different densities imposed by the EoS model. Under that light, it is interesting to consider the effect of folding nuclear theoretic calculations into the inference of the EoS.
Figure~\ref{fig:prho} shows that the \newNICER\, radius measurement does not inform the EoS below $\rho_{\rm nuc}$, something also confirmed by~\cite{Miller:2021qha}.
References~\cite{Essick:2020flb, Essick:2021kjb} further show that our GP EoS prior is designed with no strong correlations between low-density information and high-density physics by explicitly showing that the same results are obtained at high densities regardless of whether the EoS is conditioned on $\chi$EFT at low densities or not.
As such, we do not expect the \newNICER\, radius data to offer new insights about $\chi$EFT predictions within its regime of applicability, i.e.~ $\lesssim 1-2\rho_{\rm nuc}$, nor do we expect $\chi$EFT predictions to influence our conclusions about NS matter at high densities.
In contrast, the parametric EoS inference in~\cite{Raaijmakers:2021uju} is sensitive to the $\chi$EFT calculations they condition on up to $1.1\,\rhonuc$ even at the highest densities probed. Figure~7 of~\cite{Raaijmakers:2021uju} shows that the NS radii and pressures they infer with both of their parametric EoS models have some dependence on which $\chi$EFT calculation is assumed. 
This suggests that statements both about the validity of nuclear calculations based on astrophysical data and the inference of
NS properties after assuming a specific low-density calculation
must take care to avoid introducing unwanted systematic modeling assumptions through the choice of high-density EoS representation. 

In addition to $\chi$EFT and other theoretical models, several terrestrial experiments probe the EoS at densities up to $\rho_\mathrm{nuc}$.
In particular, the PREX collaboration recently measured the neutron skin thickness of lead \Rskin~\cite{Adhikari:2021phr}, which is tightly correlated with the density dependence of the nuclear symmetry energy (the difference in the energy per particle for matter that contains only neutrons and matter that contains an equal number of neutrons and protons) and therefore the pressure at $\rhonuc$~\cite{Brown:2000pd, RocaMaza:2011pm,Reed:2021nqk}.
Using a \textit{model-agnostic} nonparametric analysis similar to ours, Ref.~\cite{Essick:2021kjb} found no strong correlation between the results of several low-density experiments and high-density NS observables.
Claims to the contrary~\cite{Reed:2021nqk, Biswas:2021yge}, therefore, are driven by specific modeling assumptions, which may not be justified.
Nevertheless, the large \Rskin~reported in~\cite{Adhikari:2021phr} suggests a relatively stiff EoS at low densities, although there are other low-energy experiments (e.g.,~\cite{Roca-Maza:2015eza, Estee:2021khi,Yue:2021yfx}) and alternative interpretations of the data~\cite{Reinhard:2021utv} that favor softer EoSs.\footnote{All current bounds on the symmetry energy agree to within 2-$\sigma$.}
A stiff EoS at low densities may increase the evidence in favor of multiple stable branches, but we expect the effect to be small with current experimental uncertainties.
Given the slight tension between nuclear experiments and the fact that additional constraints at low densities will not strongly influence our conclusions from \newNICER's radius measurement, we omit nuclear experimental data from our current analysis and leave such investigations to future work.

Nevertheless, the growing number of constraints on the NS EoS is progressively sharpening our picture of dense matter. 
The radius measurement for \newNICER\, is a reminder of how different observations, experiments, and theoretical calculations complement each other by targeting different density scales inside NSs. 
Joint analyses of this ensemble of data require models for the EoS that span many orders of magnitude in pressure and density.
As a result, it is important to understand how different EoS models, both parametric and nonparametric, correlate different densities to distinguish data-driven features from those driven by the prior.
The nonparametric model we use is deliberately constructed to emphasize flexibility in EoS morphology and impose few correlations between high and low densities besides those dictated by the physical requirements of causality and thermodynamic stability.
The intra-density correlations introduced by different parametric and nonparametric EoS models will be investigated in quantitative detail in upcoming work~\cite{legred-inprep}.

\acknowledgements

Research at Perimeter Institute is supported in part by the Government of Canada through the Department of Innovation, Science and Economic Development Canada and by the Province of Ontario through the Ministry of Colleges and Universities.
S.H. is supported by the National Science Foundation, Grant PHY-1630782, and the Heising-Simons Foundation, Grant 2017-228.
P.L. is supported by National Science Foundation award PHY-1836734 and by a gift from the Dan Black Family Foundation to the Nicholas \& Lee Begovich Center for Gravitational-Wave Physics \& Astronomy.
This research has made use of data, software and/or web tools obtained from the Gravitational Wave Open Science Center (https://www.gw-openscience.org), a service of LIGO Laboratory, the LIGO Scientific Collaboration and the Virgo Collaboration.
Virgo is funded by the French Centre National de Recherche Scientifique (CNRS), the Italian Istituto Nazionale della Fisica Nucleare (INFN) and the Dutch Nikhef, with contributions by Polish and Hungarian institutes.
This material is based upon work supported by NSF's LIGO Laboratory which is a major facility fully funded by the National Science Foundation.
The authors are grateful for computational resources provided by the LIGO Laboratory and supported by National Science Foundation Grants PHY-0757058 and PHY-0823459.

\section*{Appendix: Choosing a Mass Prior}
\label{appendix}

Substantial uncertainty persists in the distribution of compact-object masses, including the question of whether the NS and BH mass distributions overlap. However, any hierarchical analysis of the EoS needs to assume a compact-object mass distribution to specify the mass prior for a given EoS.\footnote{Another common equivalent choice is to work with a prior on the central density of NSs instead of the mass, see for example~\cite{Raaijmakers:2021uju}. 
Given an EoS, there is a one-to-one mapping between the NS mass and central density; this Appendix's discussion consequently applies to these works as long as the central density distribution includes an upper and/or lower limit.}
The full framework was laid out in Sec. III B of~\cite{Landry:2020vaw} where the likelihood for an EoS model $\ep$ was written in terms of the compact-object mass distribution $P(m|\ep)$.
In the current study and~\cite{Landry:2020vaw}, we
assume a uniform mass distribution, such that the mass prior takes the general form

\begin{equation}
    P\left(m|\ep\right) = \frac{\Theta(M_\mathrm{lower} \leq m) \Theta(m\leq M_\mathrm{upper})}{M_\mathrm{upper}-M_\mathrm{lower}} \ .
\end{equation}
However, a choice still needs to be made for the lower and upper limits, 
$M_\mathrm{lower}$ and $M_\mathrm{upper}$.
We set $M_\mathrm{lower}=0.5\,\Msolar$, but $M_\mathrm{upper}$ is subject to three possible assumptions.

\subsection*{Assumption 1: The compact object might not be a NS}

The first option accounts for the possibility that the compact object in question is a BH, rather than a NS. If we do not have definite prior knowledge that it is a NS, our mass prior does not require its mass to be below the maximum NS mass.
In this case, $M_{\mathrm{upper}}$ is the maximum formation mass for the relevant type of compact object.
%
%
If $m>\Mmax(\varepsilon)$, with $\Mmax(\varepsilon)$ the TOV mass of EoS $\ep$, the radius and tidal deformability of the compact object are set to their Schwarzschild BH values, $2Gm/c^2$ and $0$, respectively. If $m \leq \Mmax(\varepsilon)$, the compact object is a NS, and its properties are set by the EoS $\varepsilon$.
The mass prior itself is independent of the EoS $\ep$ in this scenario. This is the assumption we employ for both GW170817 and GW190425.
While it is the most agnostic assumption possible, it is clearly erroneous for compact objects detected as pulsars, such as \newNICER.
  
\subsection*{Assumption 2: The astrophysical formation mechanism limits the maximum possible mass}

In the second scenario, the compact object under consideration is assumed to be a NS, but astrophysical formation mechanisms (for example, supernovae) are known \textit{a priori} not to produce NSs above a certain mass, $M_{\mathrm{pop}}$. This upper limit might be comparable to $\Mmax(\ep)$ for some EoSs.
In this case, 
\begin{equation}
    M_\mathrm{upper} = \min\left( M_\mathrm{pop}, \Mmax(\ep)\right) .
\end{equation}
%
%
While it is plausible that there may be an astrophysical limit to the mass of NSs, in practice this assumption comes at the expense of a completely arbitrary choice for $M_{\mathrm{pop}}$, given the current state of compact-object population knowledge.
As such, we do not employ it for any of the compact objects analyzed in the main body of the paper. 
Nonetheless, Fig.~\ref{fig:Mmax_comparison} compares the impact of one possible choice of $M_\mathrm{pop}$ against the other two assumptions.

\subsection*{Assumption 3: The EoS limits the maximum possible mass}


Under the third assumption, the compact object under consideration is known to be a NS, and astrophysical formation mechanisms can produce NSs as heavy as the EoS can support. 
Thus, $M_\mathrm{upper} = \Mmax(\ep)$.
%
%
In this case, the prior depends on the EoS both through the upper limit (which rejects any masses above $\Mmax$) and the normalization (which constitutes an Occam penalty against EoSs that predict masses larger than have been observed).
Stated differently, an EoS with $\Mmax(\ep)$ slightly above the most massive known pulsar will be favored compared to an EoS that predicts the existence of much more massive NSs that have not been observed.
We employ this assumption for all pulsars in our main study, including both radio and x-ray observations.
To the best of our understanding, the same assumption is employed in~\cite{Raaijmakers:2021uju,Biswas:2021yge,Miller:2021qha}.

The advantage of this assumption is that it does not rely on an explicit choice of $M_{\mathrm{pop}}$.
The disadvantage is that the lack of observations of more massive NSs is attributed to (and therefore informs) the EoS, while other factors (astrophysical conditions and selection effects) are ignored, even as potential higher-mass NS candidates have been identified~\cite{vanKerkwijk:2010mt}.
A simultaneous inference of the compact-object population and the EoS would obviate the need for choosing between Assumptions 2 and 3; instead, it would select the appropriate case as a function of the population model realization within the inference. This is possible because Assumption 3 is really just a special case of Assumption 2 in which $M_\mathrm{pop} \geq \Mmax(\ep)$ for all viable EoSs. 

\begin{figure}
    \centering
    \includegraphics[width=.49\textwidth]{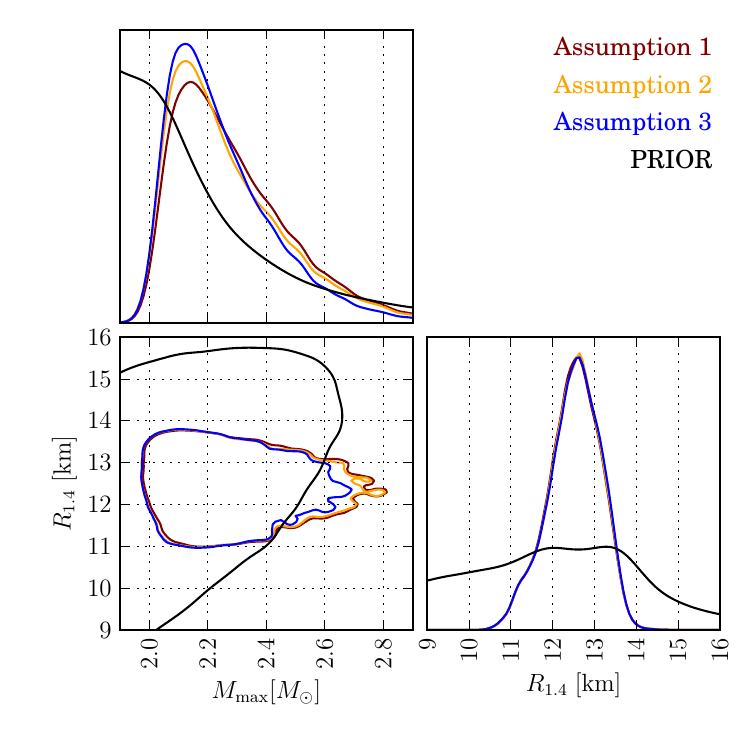}
    \caption{
        Corner plots of $\Mmax$ versus $\Rtyp$ for the three assumptions about the \newNICER\, mass prior illustrated above.
        (\textit{black}) our EoS prior.
        (\textit{brown}) Assumption 1: \newNICER\, could be either a NS or a BH.
        (\textit{orange}) Assumption 2: a hypothetical formation channel does not produce NSs with $m > 2.3\,\Msolar$, which limits the effects of the Occam penalty.
        (\textit{blue}) Assumption 3: only the EoS limits $M_\mathrm{upper}$, and EoS that support the largest masses incur the full Occam penalty.
        We see the expected ordering in the tail of the $\Mmax$ distribution; assumptions that introduce larger Occam penalties result in suppressed tails.
    }
    \label{fig:Mmax_comparison}
\end{figure}

In order to quantitatively assess the impact of assumptions about the mass prior on \newNICER, we repeat the main analysis with the alternative assumptions.
In Assumption 1, we effectively assume that \newNICER\, could be a BH.
In Assumption 2, we arbitrarily select $M_{\mathrm{pop}} = 2.3 \,\Msolar$, motivated by the approximate upper limit of the inferred \newNICER\, mass posterior~\cite{Fonseca:2021wxt}.
In Assumption 3 (same as the main body of the text), we assume $M_{\mathrm{pop}} = 3.0 \,\Msolar$, which is larger than $\Mmax$ for the vast majority of EoSs in our prior.
In Fig.~\ref{fig:Mmax_comparison}, we plot the 2-dimensional and 1-dimensional $\Mmax-\Rtyp$ marginalized prior and posterior under these three assumptions.
We find that the different mass prior choices only affect the inferred value of the maximum mass. Even then, the effect is small compared to current statistical uncertainties. Quantities determined at lower density scales, such as $\Rtyp$, are essentially unaffected.



\bibliography{references}

\end{document}